\documentclass[10pt,aps,prd,onecolumn,superscriptaddress,showpacs,nofootinbib,showkeys]{revtex4}
\pdfoutput=1

\usepackage{graphicx}
\usepackage{subfigure}
\usepackage{bm} 

\begin{document}
 
\title{Equilibration of anisotropic quark-gluon plasma produced by decays of color flux tubes}

\author{Radoslaw Ryblewski} 
\email{Radoslaw.Ryblewski@ifj.edu.pl}
\affiliation{The H. Niewodnicza\'nski Institute of Nuclear Physics, Polish Academy of Sciences, PL-31342 Krak\'ow, Poland} 

\author{Wojciech Florkowski} 
\email{Wojciech.Florkowski@ifj.edu.pl}
\affiliation{The H. Niewodnicza\'nski Institute of Nuclear Physics, Polish Academy of Sciences, PL-31342 Krak\'ow, Poland}
\affiliation{Institute of Physics, Jan Kochanowski University, PL-25406~Kielce, Poland} 

\date{\today}

\begin{abstract}
A set of kinetic equations is used to study equilibration of the anisotropic quark-gluon plasma produced by decays of color flux tubes possibly created at the very early stages of ultra-relativistic heavy-ion collisions. The decay rates of the initial color fields are given by the Schwinger formula, and the collision terms are treated in the relaxation-time approximation. By connecting the relaxation time with viscosity we are able to study production and thermalization processes in the plasma characterized by different values of the ratio of the shear viscosity to entropy density, ${\bar \eta}$. For the lowest (KSS) value of this ratio, $4\pi{\bar \eta} = 1$, and realistic initial conditions for the fields, the system approaches the viscous-hydrodynamics regime within  1--2 fm/c. On the other hand, for larger values of the viscosity, $4\pi{\bar \eta} \geq 3$, the collisions in the plasma become inefficient to destroy collective phenomena which manifest themselves as oscillations of different plasma parameters. The presence of such oscillations brings in differences between the kinetic and hydrodynamic descriptions, which suggest that the viscous-hydrodynamics approach after 1--2 fm/c is not complete if $4\pi{\bar \eta} \geq 3$ and should be extended to include dissipative phenomena connected with color conductivity.
\end{abstract}



\maketitle 

%%%%%%%%%%%%%%%%%%%%%%%%%%%%%%%%%%%%%%%%%%%%%%%%%%%%%%%%%%%
\section{Introduction}
\label{sect:intro}
%%%%%%%%%%%%%%%%%%%%%%%%%%%%%%%%%%%%%%%%%%%%%%%%%%%%%%%%%%%

In this paper we analyze equilibration of the anisotropic quark-gluon plasma produced by decays of color flux tubes possibly created at the very early stages of ultra-relativistic heavy-ion collisions \cite{Casher:1978wy,Glendenning:1983qq,
Bialas:1984wv,Bialas:1984ap,Bialas:1985is,
Gyulassy:1986jq,Bialas:1986mt,Gatoff:1987uf}. Our approach is based on the kinetic theory \cite{Elze:1986qd,Elze:1986hq,
Bialas:1987en,Dyrek:1988eb,Bialas:1989hc,
Bajan:2001fs,Florkowski:2003mm} where decay rates of the initial color fields are given by the Schwinger formula and appear as the source terms in the kinetic equations. The produced quarks and gluons interact with the mean color field and collide with each other. The collisions are described by the collision terms treated in the relaxation-time approximation \cite{Bhatnagar:1954zz,Baym:1984np,Baym:1985tna,
Heiselberg:1995sh,Wong:1996va}. By connecting the relaxation time with the system's shear viscosity, we are able to study production and thermalization processes in the quark-gluon plasma characterized by different values of the ratio of the shear viscosity to entropy density, ${\bar \eta} = \eta/\sigma$. 

In our numerical calculations we use the values of the parameter $4\pi{\bar \eta}$ which vary between $1$ and $10$. The values in the range $1 \leq 4\pi{\bar \eta} \leq 3$ have been extracted from the recent hydrodynamic analyses of ultra-relativistic heavy-ion collisions studied at RHIC and the LHC. The value $4\pi{\bar \eta}=10$ is  expected by leading log perturbative results extrapolated to RHIC and the LHC energies. On the other hand, our lowest value corresponds to the KSS (Kovtun-Son-Starinets) bound  obtained from the black hole physics \cite{Kovtun:2004de}.

The proposed model is one dimensional and assumes boost invariance, hence, it may be applied only to the early stages of the collisions and to the central rapidity region. Although similar models have been analyzed before \cite{Banerjee:1989by,Nayak:1996ex,
Nayak:1997kp,Bhalerao:1999hj}, our formulation has a few novel features. First of all, by studying different components of the energy-momentum tensor we can analyze in more detail the way how the system approaches the  hydrodynamic limit. In particular, we calculate separately the longitudinal and transverse pressure of the system and study their time dependence. Secondly, using different values of the ratio of the shear viscosity to entropy density, we may study equilibration of the quark-gluon plasma with reference to recent viscous-hydrodynamics models of ultra-relativistic heavy-ion collisions \cite{Baier:2006um,Baier:2007ix,
Romatschke:2007mq,Dusling:2007gi,Luzum:2008cw,
Song:2008hj,El:2009vj,PeraltaRamos:2010je,
Denicol:2010tr,Denicol:2010xn,
Schenke:2010rr,Schenke:2011tv,
Bozek:2009dw,Bozek:2011wa,
Niemi:2011ix,Niemi:2012ry,
Bozek:2012qs,Denicol:2012cn}.

The use of a constant value for  ${\bar \eta}$ in the kinetic equations implies that the relaxation time becomes a function of an effective temperature of the plasma. As a consequence, the present work extends earlier frameworks where the relaxation time was treated as a constant, although we do not take into account the non-abelian terms in the field equations or the minijet production, which are included in \cite{Nayak:1996ex,Nayak:1997kp} or \cite{Bhalerao:1999hj}, respectively. A general time-dependent relaxation time was used in \cite{Heiselberg:1995sh,Wong:1996va}, however, without incorporating the color fields. 

Our approach originates from the color-flux-tube model and transport theory formulated in Refs. \cite{Casher:1978wy,Glendenning:1983qq,
Bialas:1984wv,Bialas:1984ap,Bialas:1985is,
Gyulassy:1986jq,Bialas:1986mt,Gatoff:1987uf,
Elze:1986qd,Elze:1986hq,
Bialas:1987en,Dyrek:1988eb,Bialas:1989hc,
Bajan:2001fs,Florkowski:2003mm}.
At the very initial stage of a nucleus-nucleus collisions a longitudinal chromoelectric field is formed, which decays subsequently via the Schwinger mechanism. Such an initial configuration of the field is similar to the glasma state \cite{Lappi:2006fp,Romatschke:2006nk} where both chromoelectric and chromomagnetic longitudinal color fields are present at the very early stages of the collisions. We note that a similar framework to that presented in this paper has been used recently in Ref. \cite{Florkowski:2012ax} to study oscillations of the quark-gluon plasma around expanding anisotropic (color neutral) backgrounds. Our present approach goes much beyond the framework of Ref. \cite{Florkowski:2012ax} by solving the coupled kinetic and Maxwell equations in the self-consistent manner. 

The strength of the initial color fields is chosen in such a way that the effective temperature of the produced plasma reaches realistic values, $T_{\rm max} \sim 300-500$ MeV. For such initial conditions and the KSS value of the ratio of the shear viscosity to entropy density, i.e., in the case $4\pi{\bar \eta} = 1$, the plasma approaches the viscous-hydrodynamics regime within 1--2 fm/c. On the other hand, for larger values of the viscosity, $4\pi{\bar \eta} \geq 3$, the collisions in the plasma become inefficient to destroy collective phenomena which demonstrate themselves as oscillations of different plasma parameters. The presence of such oscillations brings in differences between the kinetic and viscous-hydrodynamics descriptions.

This leads us to the conclusion that after the first 1~fm/c the viscous-hydrodynamics description of the created quark-gluon plasma is quite appropriate if $4\pi{\bar \eta} = 1$, however, it is not completely satisfactory if $4\pi{\bar \eta} \geq 3$. In the latter case the hydrodynamic approach should be extended to include transport phenomena connected with color conductivity. On the other hand, it is quite interesting to observe that for $4\pi{\bar \eta} \sim 3$ the plasma might be characterized by an effective viscosity ${\bar \eta}_{\rm eff}$ which fluctuates around the fixed value of ${\bar \eta}$. Hence, in this case one may consider averaging over different color-flux-tubes which washes out the oscillations in such a way that the averaged system may be effectively well described by the viscous hydrodynamics. Quantitative analysis of this issue is left for a separate study.

The structure of the paper is as follows.  In Sec. \ref{sect:conv} we define our symbols and conventions. In Sec. \ref{sect:QGPform} the kinetic equations for the quark-gluon plasma are introduced. In Sec.~\ref{sect:boostinv} our approach is rewritten in the boost-invariant way and formal solutions of the kinetic and Maxwell equations are presented. The energy-momentum conservation laws are discussed in more detail in Sec.~\ref{sect:enmomcon}. The results of our numerical calculations are given in Sec.~\ref{sect:results}. We conclude in Sec.~\ref{sect:concl}. The details of the calculations are given in the Appendices which close the paper.

%%%%%%%%%%%%%%%%%%%%%%%%%%%%%%%%%%%%%%%%%%%%%%%%%%%%%%%%%%%
\section{Conventions}
\label{sect:conv}
%%%%%%%%%%%%%%%%%%%%%%%%%%%%%%%%%%%%%%%%%%%%%%%%%%%%%%%%%%%

We use the standard parameterizations of the four-momentum and spacetime coordinates of a particle,
\begin{eqnarray}
p^\mu &=& \left(E, {\vec p}_\perp, p_\parallel \right) =
\left(m_\perp \cosh y, p_x, p_y, m_\perp \sinh y \right), \nonumber \\
x^\mu &=& \left( t, {\vec x}_\perp, z \right) =
\left(\tau \cosh \eta, x, y, \tau \sinh \eta \right). 
\label{pandx}
\end{eqnarray} 
Here $m_\perp = \sqrt{m^2 + p_x^2 + p_y^2}$ is the transverse mass,\,\,$\tau=\sqrt{t^2 - z^2}$ is the longitudinal proper time, \,$y$ is the rapidity
\begin{eqnarray}
y = \frac{1}{2} \ln \frac{E+p_\parallel}{E-p_\parallel},  \label{y} 
\end{eqnarray}
and $\eta$ is the spacetime rapidity,
\begin{eqnarray}
\eta = \frac{1}{2} \ln \frac{t+z}{t-z}. \label{eta} 
\end{eqnarray}
The flow of matter is described by the four-vector
\begin{equation}
U^\mu = \gamma (1, v_x, v_y, v_z)\,, \quad 
\gamma=(1-v^2)^{-1}.
\label{Umu1}
\end{equation}
Throughout the paper we use natural units where $c=1$, $k_B=1$, and $\hbar=1$.

%%%%%%%%%%%%%%%%%%%%%%%%%%%%%%%%%%%%%%%%%%%%%%%%%%%%%%%%%%%
\section{QGP formation by decays of color flux tubes}
\label{sect:QGPform}
%%%%%%%%%%%%%%%%%%%%%%%%%%%%%%%%%%%%%%%%%%%%%%%%%%%%%%%%%%%

%%%%%%%%%%%%%%%%%%%%%%%%%%%%%%%%%%%%%%%%%%%%%%%%%%%%%%%%%%%
\subsection{Kinetic equations, color isotopic charge and color hypercharge}
\label{sect:color}
%%%%%%%%%%%%%%%%%%%%%%%%%%%%%%%%%%%%%%%%%%%%%%%%%%%%%%%%%%%

Our approach is based on the abelian dominance approximation in which the kinetic equations for quarks, antiquarks, and charged gluons can be written as follows \cite{Elze:1986qd,Elze:1986hq,Bialas:1987en} 
\begin{equation}
\left( p^\mu \partial_\mu + g{\mbox{\boldmath $\epsilon$}}_i \cdot 
{\bf F}_{}^{\mu \nu } p_\nu \partial_\mu^p\right) G_{if}(x,p) = \frac{dN_{if}}{d\Gamma_{\rm inv} }+ C_{if},  
\label{kineq}
\end{equation}
\begin{equation}
\left( p^\mu \partial_\mu - g{\mbox{\boldmath $\epsilon$}}_i \cdot 
{\bf F}_{}^{\mu \nu } p_\nu \partial_\mu^p\right) \bar{G}_{if}(x,p) = \frac{dN_{if}}{d\Gamma_{\rm inv} } + \bar{C}_{if},  
\label{kineaq}
\end{equation}
\begin{equation}
\left( p^\mu \partial _\mu + g{\mbox{\boldmath $\eta$}}_{ij} \cdot 
{\bf F}_{}^{\mu \nu } p_\nu \partial_\mu^p\right) \widetilde{{G}}_{ij}(x,p) = 
\frac{d\widetilde{N}_{ij}}{d\Gamma_{\rm inv} }+ \tilde{C}_{ij}. \label{kineg}
\end{equation}
Here $G_{if}(x,p)\ $, $\bar{G}_{if}(x,p)$ and ${G}_{ij}(x,p)$ are the phase-space densities of quarks, antiquarks and gluons, respectively. The indices $i,j=(1,2,3)$ define the color of quarks and gluons, while the index $f=(u,d,s,...)$ defines the quark flavor. In the definitions of the gluon distribution functions $\widetilde{{G}}_{ij}(x,p)$ and other similar objects the two color indices $i$ and $j$ must be different.

The terms on the left-hand-sides of Eqs.~(\ref{kineq})--(\ref{kineg}) describe the free motion of particles and their interaction with the mean field ${\bf F}_{\mu \nu }$. In this work, the only non-zero components of the tensor ${\bf F}^{\mu \nu }=(F^{\mu \nu }_{(3)},F^{\mu \nu }_{(8)})$ are those corresponding to the longitudinal chromoelectric  field ${\mbox{\boldmath $\cal E$}} = (F^{30}_{(3)},F^{30}_{(8)})$. The quarks couple to the chromoelectric field ${\mbox{\boldmath $\cal E$}}$ through the charges \cite{HuangQL}
\begin{equation}
\mbox{\boldmath $\epsilon$}_{1} = \frac{1}{2}\left(\! 1,\sqrt{\frac{1}{3}}\right) \!, 
\mbox{\boldmath $\epsilon$}_{2} = \frac{1}{2}\left(\! -1,\sqrt{\frac{1}{3}}\right) \!, 
\mbox{\boldmath $\epsilon$}_{3} = \left(\! 0,-\sqrt{\frac{1}{3}}\right) .
\label{qcharge}
\end{equation}
The charged gluons couple to ${\mbox{\boldmath $\cal E$}}$ through the charges ${\mbox{\boldmath $\eta $}}_{ij}$ defined by the relation 
\begin{equation}
{\mbox{\boldmath $\eta$}}_{ij}={\mbox{\boldmath $\epsilon$}}_{i}-{%
\mbox{\boldmath $\epsilon$}}_{j}.  \label{gcharge}
\end{equation}
The two components of the color charges (\ref{qcharge}) and (\ref{gcharge}) are called the color isotopic charge and color hypercharge \cite{HuangQL}.

%%%%%%%%%%%%%%%%%%%%%%%%%%%%%%%%%%%%%%%%%%%%%%%%%%%%%%%%%%%
\subsection{Initial conditions}
\label{sect:init}
%%%%%%%%%%%%%%%%%%%%%%%%%%%%%%%%%%%%%%%%%%%%%%%%%%%%%%%%%%%

The Gauss law applied to a color flux tube yields ${\mbox{\boldmath $\cal E$}} {\cal A} = k g {\bf q}$, where ${\cal A} = \pi r^2$ denotes the area of the transverse cross section of the tube, $k$ is the number of color charges at the end of the tube and $g{\bf q} = g (q_{(3)},q_{(8)})$ is the color charge of a quark or a gluon. Since the string tension is the energy of an elementary tube (with $k=1$) per unit length we may write
\begin{equation}
\sigma =\frac 12 {\cal A} \, {\mbox{\boldmath $\cal E$}}\cdot {\mbox{\boldmath $\cal E$}}
=\frac{g^2}{2 {\cal A}} \, {\bf q\cdot q.}  \label{strten}
\end{equation}
Substitution of the quark charges (\ref{qcharge}) into Eq.~(\ref{strten}) gives (independently of the color index $i$) $\sigma _q=g^2/(6 {\cal A})$.
Similarly, the gluon charges (\ref{gcharge}) give (independently of the color indices $i,j$) $\sigma _g=g^2/(2 {\cal A})$. We conclude that the string tension of a tube spanned by gluons is three times larger than the string tension of a quark tube. The Gauss law can be rewritten in the following form 
\begin{equation}
{\mbox{\boldmath $\cal E$}}=\sqrt{\frac{2\sigma _g}{\pi r^2}}k{\bf q}
=\sqrt{\frac{6\sigma _q}{\pi r^2}}k{\bf q.}  \label{inite}
\end{equation}
This equation determines the value of the initial chromoelectric field spanned by the two receding nuclei.

We have three parameters which characterize an elementary tube: the string tension $\sigma $, the strong coupling constant $g$, and the tube radius $r$. For the standard value $\sigma_q=1$ GeV/fm ($\sigma_g=3$ GeV/fm) which is used in our calculations we find the following relation between $g$ and $r$,
\begin{equation}
g^2 = 6\, {\cal A} \, \hbox{GeV/fm} \approx 30 \, \pi r^2/\hbox{fm}^2.  \label{coupl}
\end{equation}
For a single tube we assume $\pi r^2=$ 1 fm$^2$, hence, we find $g=5.48$. Consequently, our coupling constant is quite large, which excludes any perturbative treatment of the considered physical processes. The number of color charges $k$ may be obtained from the hypothesis of random walk in color space \cite{Biro:1984cf}
\[
k=\sqrt{\hbox{number of collisions}}=\sqrt{\frac{d\nu }{d^2s}\pi r^2}, 
\]
where $d\nu /d^2s$ is the number of collisions per unit transverse area. In the original papers on the color-flux-tube model the values of $k$ between 1 and 5 were used. In this paper, to obtain the initial effective temperature of the plasma which is similar to that used in the hydrodynamic calculations describing the RHIC and the LHC data, we use $k=5$ and $k=10$.

%%%%%%%%%%%%%%%%%%%%%%%%%%%%%%%%%%%%%%%%%%%%%%%%%%%%%%%%%%%
\subsection{Schwinger tunneling}
\label{sect:schwinger}
%%%%%%%%%%%%%%%%%%%%%%%%%%%%%%%%%%%%%%%%%%%%%%%%%%%%%%%%%%%

The terms $dN/d\Gamma_{\rm inv}$ on the right-hand-sides of Eqs.~(\ref{kineq})--(\ref{kineg}) describe  production of quarks and gluons due to the decay of the chromoelectric field. In the reference frame where the particles emerge from the vacuum with the vanishing longitudinal momentum, the production rate of quarks in the chromoelectric field is given by the formula
\begin{equation}
\frac{dN_{if}}{d\Gamma_{\rm inv} }=\frac{\Lambda _i}{4\pi ^3}\left| \ln \left( 1-\exp
\left( -\frac{\pi m_{f\perp }^2}{\Lambda _i}\right) \right) \right| \delta
\left( p_{\Vert }\right) p^0\equiv {\cal R}_{if}\ \delta \left( p_{\Vert
}\right) p^0,  \label{qrate}
\end{equation}
where $m_{f\perp} = \sqrt{m_f^2 + p_x^2 + p_y^2}$ is the transverse mass,
\begin{equation}
\Lambda _i=\left( g\left| {\mbox{\boldmath $\epsilon$}}_i\cdot {%
\mbox{\boldmath $\cal E$}}\right| -\sigma _q\right) \theta \left( g\left| {%
\mbox{\boldmath $\epsilon$}}_i\cdot {\mbox{\boldmath $\cal E$}}\right|
-\sigma _q\right),  \label{Lami}
\end{equation}
and $\theta$ is the step function. The quantity $\Lambda _i$ describes the effective force acting on the tunneling quarks. The effect of the screening of the initial field by the tunneling particles is taken into account by subtraction of the ``elementary force'' characterized by the quark string tension $\sigma_q$ \cite{Glendenning:1983qq,%
Gyulassy:1986jq}. Similarly, for gluons one can find 
\begin{equation}
\frac{d\widetilde{N}_{ij}}{d\Gamma_{\rm inv} }=\frac{\Lambda _{ij}}{4\pi ^3}\left| \ln
\left( 1+\exp \left( -\frac{\pi p_{\bot }^2}{\Lambda _{ij}}\right) \right)
\right| \delta \left( p_{\Vert }\right) p^0\equiv \widetilde{{\cal R}}_{ij}\
\delta \left( p_{\Vert }\right) p^0,  \label{grate}
\end{equation}
where the effective force is 
\begin{equation}
\Lambda _{ij}=\left( g\left| {\mbox{\boldmath $\eta$}}_{ij}\cdot {%
\mbox{\boldmath $\cal E$}}\right| -\sigma _g\right) \theta \left( g\left| {%
\mbox{\boldmath $\eta$}}_{ij}\cdot {\mbox{\boldmath $\cal E$}}\right|
-\sigma _g\right).   \label{Lamij}
\end{equation}
The covariant form of the production rates which is valid in arbitrary reference frames is given in  Sec.~\ref{sect:boostinvkineq}.

%%%%%%%%%%%%%%%%%%%%%%%%%%%%%%%%%%%%%%%%%%%%%%%%%%%%%%%%%%%
\subsection{Collision terms and Landau matching}
\label{sect:col-terms}
%%%%%%%%%%%%%%%%%%%%%%%%%%%%%%%%%%%%%%%%%%%%%%%%%%%%%%%%%%%

The terms $C$ are the collision terms which we treat in the relaxation time approximation \cite{Bhatnagar:1954zz,Baym:1984np,Baym:1985tna,
Heiselberg:1995sh,Wong:1996va},
\begin{eqnarray}
C_{if} = p \cdot U \, \frac{G^{\rm eq}_{if}-G_{if}}{\tau_{\rm eq}}, \quad
\bar{C}_{if} = p \cdot U \, \frac{\bar{G}^{\rm eq}_{if}-\bar{G}_{if}}{\tau_{\rm eq}},
\quad
\tilde{C}_{ij} = p \cdot U \, \frac{\tilde{G}^{\rm eq}_{ij}-\tilde{G}_{ij}}{\tau_{\rm eq}}.
\label{col-terms}
\end{eqnarray}
The background equilibrium distributions $G^{\rm eq}_{if}$, $\bar{G}^{\rm eq}_{if}$, and $\tilde{G}^{\rm eq}_{ij}$ are all equal and given by the Boltzmann distribution~\footnote{The generalization of our approach in such a way as to include quantum Bose-Einstein or Fermi-Dirac statistics is quite straightforward. It typically brings only different normalization factors.}
\begin{eqnarray}
G^{\rm eq} = \frac{2}{(2\pi)^3} 
\exp\left(-\frac{p \cdot U}{T} \right).
\label{Boltzmann}
\end{eqnarray}
Similarly, we assume that all the relaxation times are equal. The effect of having different relaxation times for different particle species has been recently analyzed in a similar framework in Ref.~\cite{Florkowski:2012as}.

The factor 2 in Eq.~(\ref{Boltzmann}) accounts for spin degeneracy. The effective temperature $T$ is obtained from the Landau matching 
\begin{eqnarray}
\int dP \, p \cdot U \, p^\nu \left(
\sum_{i=1}^3 \sum_f (G^{\rm eq}_{if}-G_{if}
+ \bar{G}^{\rm eq}_{if}-\bar{G}_{if})
+ \sum_{i,j=1}^3 (\tilde{G}^{\rm eq}_{ij}-\tilde{G}_{ij}) \right) =0
\label{LM0}
\end{eqnarray}
or
\begin{eqnarray}
\int dP \, p \cdot U \, p^\nu \left(
(6 N_f + 6) G^{\rm eq}- \sum_i \sum_f (G_{if} +\bar{G}_{if}) - \sum_{i,j=1}^3
\tilde{G}_{ij} \right) =0. \label{LM}
\end{eqnarray}
The factors 6 account for six types of quarks (three quarks and three antiquarks) and six types of gluons (the two extra gluon degrees of freedom are treated as fields). As we shall see below, the Landau matching is reduced to the condition which demands that the energy density obtained from the quark and gluon distribution functions is equal to the energy density determined from the equilibrium background (defined by the sum of thermal distributions).

%%%%%%%%%%%%%%%%%%%%%%%%%%%%%%%%%%%%%%%%%%%%%%%%%%%%%%%%%%%
\subsection{Relaxation time}
\label{sect:taueq}
%%%%%%%%%%%%%%%%%%%%%%%%%%%%%%%%%%%%%%%%%%%%%%%%%%%%%%%%%%%

As stated above, the relaxation time used in Eq.~(\ref{col-terms}) is the same for all parton species and independent of momentum. On the other hand, it may depend on the proper time. In the numerical calculations we use the following relation between the relaxation time and the viscosity  \cite{Anderson:1974,Czyz:1986mr,Dyrek:1986vv,
cerc,Romatschke:2011qp,Florkowski:2013lza,Florkowski:2013lya}.
\begin{eqnarray}
\tau_{\rm eq}(\tau) = \frac{5 {\bar \eta}}{T(\tau)}.
\label{taueq}
\end{eqnarray}
Here ${\bar \eta}$ is the ratio of the viscosity to the entropy ratio which is treated as a constant in our approach. We consider three values of ${\bar \eta}$: 
\begin{eqnarray}
{\bar \eta} = \frac{1}{4\pi}\,,\, \frac{3}{4\pi}\,,  \frac{10}{4\pi}.
\label{etabars}
\end{eqnarray}
The first two numbers on the right-hand-side of Eq.~(\ref{etabars}) determine the viscosity range extracted from the recent hydrodynamic analyses of relativistic heavy-ion collisions studied at RHIC and the LHC. The last value is on the order expected by leading log perturbative results extrapolated to RHIC and the LHC energies.

%%%%%%%%%%%%%%%%%%%%%%%%%%%%%%%%%%%%%%%%%%%%%%%%%%%%%%%%%%%
\subsection{Maxwell equations}
\label{sect:Maxwell}
%%%%%%%%%%%%%%%%%%%%%%%%%%%%%%%%%%%%%%%%%%%%%%%%%%%%%%%%%%%

Equations (\ref{kineq})--(\ref{kineg}) determine the time changes of parton densities caused by the presence of the mean field and parton collisions. In order to obtain a self-consistent set of equations we should have also the dynamic equation for the field. It can be written in the following Maxwell form
\begin{equation}
\partial _\mu {\bf F}^{\mu \nu }(x)={\bf j}^\nu (x)+{\bf j}_D^\nu (x),
\label{Maxwell}
\end{equation}
where
\begin{eqnarray}
{\bf j}^\nu (x) &=& g\int dP \, p^\nu \left[ \sum_i{\mbox{\boldmath $\epsilon$}}_i 
\sum_f\left( G_{if}(x,p) - \bar{G}_{if}(x,p)\right) 
+\sum_{i,j=1}^3{\mbox{\boldmath $\eta$}}_{ij}\widetilde{G}_{ij}(x,p)\right] 
\equiv  g\int dP \, p^\nu A(\tau,w,p_\perp)
\label{condcur}
\end{eqnarray}
and 
\begin{eqnarray}
{\bf j}_D^\nu (x)&=&\int dP\left[ \sum_i\sum_f\frac{dN_{if}}{d\Gamma_{\rm inv} }{\bf d}_{if}^\nu 
+ \sum_{i>j}\frac{d\widetilde{N}_{ij}}{d\Gamma_{\rm inv} }{\bf d}_{ij}^\nu \right] .  \label{convcur}
\end{eqnarray}
Here $dP = d^3p/E$ is the invariant momentum integration measure~\footnote{The spin degeneracy factor of 2 is included in the production rates (\ref{qrate}) and (\ref{grate}). Therefore,  we include it also in the definitions of the quark and gluon distribution functions.}. The current (\ref{Maxwell}) has two components. The first one (conductive current), defined by Eq.~(\ref{condcur}), is related to the simple fact that particles carry color charges ${\mbox{\boldmath $\epsilon $}}_i$ and ${\mbox{\boldmath $\eta $}}_{ij}$. The second one (displacement current), defined by Eq.~(\ref{convcur}),  has the origin in the tunneling of quarks and gluons, hence, in the creation of color charges from the vacuum. On the right-hand-side of Eq.~(\ref{condcur}) we have introduced $A(\tau,w,p_\perp)$ which is an antisymmetric function of $w$.

The quantities ${\bf d}^\nu $ are the dipole moments of the produced pairs whose third Lorentz components equal
\begin{equation}
{\bf d}_{if}^3 = 
\frac{2{\mbox{\boldmath $\epsilon$}}_i}
{\ {\mbox{\boldmath $\epsilon$}}_i \cdot {\mbox{\boldmath $\cal E$}}}
\sqrt{p_{\bot }^2+m_f^2}, \quad
{\bf d}_{ij}^3 = 
\frac{2{\mbox{\boldmath $\eta$}}_{ij}}
{{\mbox{\boldmath $\eta$}}_{ij} \cdot {\mbox{\boldmath $\cal E$}}}
p_{\bot }.  \label{dipols}
\end{equation}
Below, in Sec. \ref{sect:EOM}, we show that the form of Eqs.~(\ref{Maxwell})--(\ref{dipols}) follows directly from the energy-momentum conservation law.

%%%%%%%%%%%%%%%%%%%%%%%%%%%%%%%%%%%%%%%%%%%%%%%%%%%%%%%%%%%
\section{Implementation of boost invariance}
\label{sect:boostinv}
%%%%%%%%%%%%%%%%%%%%%%%%%%%%%%%%%%%%%%%%%%%%%%%%%%%%%%%%%%%

%%%%%%%%%%%%%%%%%%%%%%%%%%%%%%%%%%%%%%%%%%%%%%%%%%%%%%%%%%%
\subsection{Boost-invariant variables}
\label{sect:boostinvvar}
%%%%%%%%%%%%%%%%%%%%%%%%%%%%%%%%%%%%%%%%%%%%%%%%%%%%%%%%%%%

In the case of one-dimensional boost-invariant expansion, all scalar functions of time and space coordinates (for example: energy density, transverse and longitudinal pressure, temperature) should depend only on the proper time $\tau$. In addition, the hydrodynamic flow $U^\mu$ should have the form (\ref{Umu1}) with $v_x=v_y=0$ and $v_z=z/t$, hence $U^\mu = \left(t/\tau,0,0,z/\tau\right)$ \cite{Bjorken:1982qr}.

The phase-space distribution functions behave also like scalars under Lorentz transformations. The requirement of boost invariance implies in this case that they may depend only on the variables $\tau$, $w$ and $\vec{p}_\perp$ \cite{Bialas:1984wv}. The boost-invariant variable $w$ is defined by the formula
\begin{equation}
w =  t p_\parallel - z E.
\label{binvv1}
\end{equation}
We note that at $z=0$ the variable $w$ is reduced to the longitudinal momentum multiplied by the time coordinate $t$. Knowing $w$ and $\vec{p}_\perp$ we define 
\begin{equation}
v = Et-p_\parallel z = \tau \, p \cdot U  =
\sqrt{w^2+\left( m^2+\vec{p}_\perp^{\,\,2}\right) \tau^2}.  
\label{binvv2}
\end{equation}
At $z=0$, $v$ is reduced to the energy multiplied by $t$. In the numerical calculations we neglect the quark masses and set $m=0$. From (\ref{binvv1}) and (\ref{binvv2}) one can easily find the energy and the longitudinal momentum of a parton,
\begin{equation}
E= \frac{vt+wz}{\tau^2},\quad p_{\Vert }=\frac{wt+vz}{\tau^2}.  
\label{binvv3}
\end{equation}
The integration measure in the momentum sector of the phase-space is 
\begin{equation}
dP = 2 \, d^4p \, \delta \left( p^2\right) \theta (p^0)
=\frac{dp_{\Vert }}{p^0}d^2p_{\bot }
=\frac{dw}vd^2p_{\bot }.  \label{binvm}
\end{equation}

A chromoelectric field ${\mbox{\boldmath $\cal E$}} = {\bf F}^{30}$ does not change under Lorentz transformations along the $z$-axis, thus it may be also treated as a scalar and written in the form
\begin{eqnarray}
\nonumber \\
{\mbox{\boldmath $\cal E$}(\tau)} &=& - 2 \, \frac{d{\bf h}(\tau)}{du}
= - \frac 1\tau \frac{d{\bf h}(\tau)}{d\tau },  
\label{binve} \\ \nonumber 
\end{eqnarray}
where ${\bf h}$ is a function of the variable $\tau$ only.

%%%%%%%%%%%%%%%%%%%%%%%%%%%%%%%%%%%%%%%%%%%%%%%%%%%%%%%%%%%
\subsection{Boost-invariant form of the kinetic equations and their solutions}
\label{sect:boostinvkineq}
%%%%%%%%%%%%%%%%%%%%%%%%%%%%%%%%%%%%%%%%%%%%%%%%%%%%%%%%%%%

Using the boost-invariant variables introduced in the previous Section we find simple forms of the terms appearing in the kinetic equations (\ref{kineq})--(\ref{kineg}), namely
\begin{eqnarray}
p^\mu \partial_\mu G = \frac{v}{\tau} \frac{\partial G}{\partial \tau}, \quad
{\bf F}_{}^{\mu \nu } p_\nu 
\partial_\mu^p  G  = 
{\mbox{\boldmath $\cal E$}} v 
\frac{\partial G}{\partial w}, \quad
\frac{dN}{d\Gamma_{\rm inv}}
= v \delta(w) \cal{R}.
\label{binvterms}
\end{eqnarray}
Using (\ref{binvterms}) in (\ref{kineq})--(\ref{kineg}) one obtains
\begin{eqnarray}
\left( \frac{\partial}{\partial \tau}
- \frac{dh_i}{d\tau} \frac{\partial}{\partial w} \right) G_{if} &=& \tau {\cal R}_{if} \delta(w) + \frac{G^{\rm eq}-G_{if}}{\tau_{\rm eq}}, \nonumber \\
\left( \frac{\partial}{\partial \tau}
+ \frac{dh_i}{d\tau} \frac{\partial}{\partial w} \right) \bar{G}_{if} &=& \tau {\cal R}_{if} \delta(w) + \frac{G^{\rm eq}-\bar{G}_{if}}{\tau_{\rm eq}}, 
\label{kineqs} \\
\left( \frac{\partial}{\partial \tau}
- \frac{dh_{ij}}{d\tau} \frac{\partial}{\partial w} \right) \tilde{G}_{ij} &=& \tau {\tilde R}_{ij} \delta(w) + \frac{G^{\rm eq}-\tilde{G}_{ij}}{\tau_{\rm eq}}.
\nonumber
\end{eqnarray}
Here we have introduced the functions
\begin{eqnarray}
h_i \left( \tau \right) 
= g{\mbox{\boldmath $\epsilon$}}_i \cdot 
 {\bf h}\left( \tau \right),  
\quad
h_{ij}\left( \tau \right) 
= g{\mbox{\boldmath $\eta$}}_{ij} \cdot 
 {\bf h}\left(\tau \right) .  
\label{hs} 
\end{eqnarray}
One may notice that the distribution functions in (\ref{kineqs}) satisfy the following symmetry relations 
\begin{eqnarray}
\bar{G}_{if} 
\left( \tau ,w,p_\perp \right) &=& 
G_{if}\left( \tau,-w,p_{\bot }\right) ,
\nonumber \\
\tilde{G}_{ij}\left( \tau ,w,p_{\bot}\right) &=&
\tilde{G}_{ji}\left( \tau ,-w,p_{\bot }\right) .
\label{symofg} 
\end{eqnarray}
Equations (\ref{kineqs}) have general solutions of the form
\begin{eqnarray}
G_{if}\left(\tau,w,p_{\bot }\right)
&=& \int\limits_0^\tau d\tau^\prime D(\tau,\tau^\prime) \left[ \tau^\prime 
{\cal R}_{if}(\tau^\prime,p_\perp) 
\delta(\Delta h_i + w) + \frac{G_{\rm eq}(\tau^\prime,\Delta h_i + w,p_\perp)}{\tau_{\rm eq}(\tau^\prime)}
\right], \nonumber \\
{\bar G}_{if}\left(\tau,w,p_{\bot }\right)
&=& \int\limits_0^\tau d\tau^\prime D(\tau,\tau^\prime) \left[ \tau^\prime 
{\cal R}_{if}(\tau^\prime,p_\perp) 
\delta(\Delta h_i - w) 
+ \frac{{G}_{\rm eq}(\tau^\prime,\Delta h_i - w,p_\perp)}{\tau_{\rm eq}(\tau^\prime)}
\right], \label{formsol} \\
G_{ij}\left(\tau,w,p_{\bot }\right)
&=& \int\limits_0^\tau d\tau^\prime D(\tau,\tau^\prime) \left[ \tau^\prime 
{\cal R}_{ij}(\tau^\prime,p_\perp) 
\delta(\Delta h_{ij} + w) 
+ \frac{G_{\rm eq}(\tau^\prime,\Delta h_{ij} 
+ w,p_\perp)}{\tau_{\rm eq}(\tau^\prime)}
\right]. \nonumber 
\end{eqnarray}
Here we have introduced the damping function
\begin{eqnarray}
D(\tau_2,\tau_1) = \exp\left[-\int\limits_{\tau_1}^{\tau_2}
\frac{d\tau^{\prime\prime}}{\tau_{\rm eq}(\tau^{\prime\prime})} \right],
\end{eqnarray}
and the functions $\Delta h_i\left( \tau ,\tau ^{\prime }\right) = h_i(\tau) - h_i(\tau^\prime)$ and $\Delta h_{ij}\left( \tau ,\tau ^{\prime }\right) = h_{ij}(\tau) - h_{ij}(\tau^\prime)$. Equations (\ref{formsol}) are generalizations of the formulas used previously in the literature where the relaxation time was constant \cite{Banerjee:1989by}. Using the boost-invariant variables, the equilibrium distribution function may be rewritten in the following form
\begin{eqnarray}
G^{\rm eq}(\tau,w,p_\perp) =
\frac{2}{(2\pi)^3} \exp\left[
- \frac{\sqrt{w^2+p_\perp^2 \tau^2}}{T(\tau) \tau}  \right].
\end{eqnarray}
%

%%%%%%%%%%%%%%%%%%%%%%%%%%%%%%%%%%%%%%%%%%%%%%%%%%%%%%%%%%%
\subsection{Boost-invariant form of the conductive and displacement currents}
\label{sect:binvcurrents}
%%%%%%%%%%%%%%%%%%%%%%%%%%%%%%%%%%%%%%%%%%%%%%%%%%%%%%%%%%%

\subsubsection{Conductive current}

By the explicit calculations starting with Eq.~(\ref{condcur}) one can check that the boost-invariant conductive current has the form
\begin{eqnarray}
{\bf j}^\nu = \left(z,0,0,t\right) {\bf J}(\tau),
\label{bfj}
\end{eqnarray}
where 
\begin{eqnarray}
{\bf J}(\tau) = \frac{g}{\tau^2} \int \frac{dw}{v} d^2p_\perp w \, A(\tau,w,p_\perp).
\end{eqnarray}
The function $A(\tau,w,p_\perp)$ is a combination of the distribution functions multiplied by the appropriate color charges, see Eq.~(\ref{condcur}). This leads us to the expression
\begin{eqnarray}
{\bf J}(\tau) &=& \frac{g N_f}{4\pi^3} \, \sum_i {\mbox{\boldmath $\epsilon$}}_i 
\int_0^\tau d\tau^\prime D(\tau,\tau^\prime)
\left[2 \tau^\prime \Lambda_i^2(\tau^\prime)
\beta_i(\tau,\tau^\prime) C^-(\beta_i(\tau,\tau^\prime)) - \Sigma^J_i(\tau,\tau^\prime) \right]
\nonumber \\
&& +\frac{g}{4\pi^3} \, \sum_{i>j} {\mbox{\boldmath $\eta$}}_{ij} 
\int_0^\tau d\tau^\prime D(\tau,\tau^\prime)
\left[2 \tau^\prime \Lambda_{ij}^2(\tau^\prime)
\beta_{ij}(\tau,\tau^\prime) C^+(\beta_{ij}(\tau,\tau^\prime)) - \Sigma^J_{ij}(\tau,\tau^\prime) \right]. \label{bfJ}
\end{eqnarray}
The function $\beta(\tau,\tau^\prime)$ is defined by the formula (below, for clarity of notation, in most places we skip the color indices $i,j$)
\begin{eqnarray}
\beta(\tau,\tau^\prime) = \frac{\Delta h(\tau,\tau^\prime)}{\sqrt{\Delta h^2(\tau,\tau^\prime)+\Lambda(\tau^\prime) \tau^2/\pi}}, 
\label{beta}
\end{eqnarray}
whereas the functions $C^\pm$ are defined as integrals \cite{Bialas:1987en}
\begin{eqnarray}
C^{\pm}(\beta) = \int_0^\infty  d\xi\,
\frac{|\ln\left( 1\pm e^{-\xi}\right)|}{\sqrt{\beta^2+(1-\beta^2)\xi}}.
\label{Cpm}
\end{eqnarray}
The function $\Sigma^J(\tau,\tau^\prime)$ appearing in (\ref{bfJ}) is defined as an integral over the equilibrium distribution function
\begin{eqnarray}
\Sigma^J(\tau,\tau^\prime) = \frac{8\pi^3}{\tau_{\rm eq}(\tau^\prime)} \int dw
\int d^2p_\perp \, \frac{w}{v} \, G^{\rm eq}(\tau^\prime,\Delta h(\tau,\tau^\prime)+w, p_\perp).
\label{SigmaJ}
\end{eqnarray}
In the limit $\tau_{\rm eq} \longrightarrow \infty$ the $\Sigma^J$ terms in (\ref{bfJ}) vanish and Eq.~(\ref{bfJ}) is reduced to the expression derived for the first time in \cite{Bialas:1987en}.

\subsubsection{Displacement current}

The boost-invariant displacement current has the same form as the conductive current, namely
\begin{eqnarray}
{\bf j}_D^\nu = \left(z,0,0,t\right)
 {\bf J}_D(\tau),
\label{bfjd}
\end{eqnarray}
where 
\begin{eqnarray}
%\nonumber \\
{\bf J}_D\left( \tau \right) &=&
\frac{g N_f}{2\pi ^3\tau}
\sum_i
\frac{{\mbox{\boldmath $\epsilon$}}_i 
\Lambda _i\left(\tau \right)}
{\Lambda _i\left(\tau \right)+\sigma_q}
\sqrt{\frac{\Lambda _i\left(\tau \right)}{\pi}}
\hbox{sgn}({\mbox{\boldmath
$\epsilon$}}_{i}  \cdot 
{\mbox{\boldmath $\cal E$}}) D^{-}(0)
 \nonumber \\
&&+\frac{g}{2\pi ^3\tau}
\sum_{i>j}
\frac{{\mbox{\boldmath $\eta$}}_{ij} 
\Lambda_{ij}\left(\tau \right)}
{\Lambda_{ij}\left(\tau \right)+\sigma_g}
\sqrt{\frac{\Lambda_{ij}\left(\tau \right)}{\pi}}
\hbox{sgn}({\mbox{\boldmath
$\eta$}}_{ij}  \cdot 
{\mbox{\boldmath $\cal E$}}) D^{+}(0). \label{jbfd}
\end{eqnarray}
and the functions $D^\pm(\beta)$ are defined as integrals \cite{Bialas:1987en}
\begin{eqnarray}
D^{\pm}(\beta) = \int_0^\infty d\xi\,
\sqrt{\beta^2+(1-\beta^2)\xi} \,\,
|\ln\left( 1\pm e^{-\xi}\right)|.
\label{Dpm}
\end{eqnarray}
We note that Eq.~(\ref{jbfd}) agrees again with the expression used before in \cite{Bialas:1987en}.

%%%%%%%%%%%%%%%%%%%%%%%%%%%%%%%%%%%%%%%%%%%%%%%%%%%%%%%%%%%%%%%%%%%%%%%%%%%
\subsection{Boost-invariant Maxwell equations}
%%%%%%%%%%%%%%%%%%%%%%%%%%%%%%%%%%%%%%%%%%%%%%%%%%%%%%%%%%%%%%%%%%%%%%%%%%%

With the all substitutions required by the boost invariance, the field equation (\ref{Maxwell}) may be written in a very compact form as 
\begin{equation}
\frac{d^2{\bf h}\left( \tau \right) }{d\tau ^2}=\frac 1\tau \frac{d{\bf h}%
\left( \tau \right) }{d\tau ^{}}+\tau ^2\left[ {\bf J}
\left( \tau \right) +{\bf J}_D\left( \tau \right) \right].  
\label{biMaxwell}
\end{equation}
This is an integro-differential equation for the function ${\bf h}\left(\tau \right)$ because the conductive current ${\mbox{\boldmath $\bf J$}} \left( \tau \right) $ depends not only on ${\bf h}\left( \tau \right)$ but also on the values of ${\bf h(}\tau ^{\prime })$ for \mbox{$0\leq \tau^{\prime }\leq \tau $}. Equation (\ref{biMaxwell}) has to be solved numerically for given initial values. These are taken in the form 
\begin{equation}
{\bf h}\left( 0\right) =0,\quad \frac 1\tau \frac{d{\bf h}}{d\tau }(0)=-{%
\mbox{\boldmath $\cal E$}}_0{\,=}-\sqrt{\frac{6\sigma_q}{\pi r^2}}k{\bf q}.
\label{initcon}
\end{equation}
In practice, to solve Eq.~(\ref{biMaxwell}) we apply the iterative method introduced in \cite{Banerjee:1989by}. We first assume a certain temperature time profile in the background thermal distributions, i.e., we start with an arbitrary  function $T(\tau)$ and solve the equation for the field ${\bf h}(\tau)$ in this background. At the same time we determine the new temperature profile from the Landau matching condition. In the next step, we use the new temperature profile to solve the field equations and determine the next temperature profile from the Landau matching condition. Repeating this procedure several times, we come to the stable solution for $T(\tau)$ and the field ${\bf h}(\tau)$.

The solution of (\ref{biMaxwell}) is independent of the initial condition for ${\bf h}\left( \tau \right) $ because of the cancellations connected with the gauge transformation which leaves ${\mbox{\boldmath $\cal E$}}$ unchanged. Since the exchange of color charges at the initial stage of a heavy-ion collision leads to the color fields spanned by gluons, we assume that ${\bf q}$ is one of the gluon color charges ${\mbox{\boldmath $\eta $}}_{ij}$. In reality, after a collision the color distribution of nuclear discs may be strongly fluctuating in the transverse direction. In this approach such fluctuations are smoothed out.

%%%%%%%%%%%%%%%%%%%%%%%%%%%%%%%%%%%%%%%%%%%%%%%%%%%%%%%%%%%
\section{Energy-momentum conservation law}
\label{sect:enmomcon}
%%%%%%%%%%%%%%%%%%%%%%%%%%%%%%%%%%%%%%%%%%%%%%%%%%%%%%%%%%%

%%%%%%%%%%%%%%%%%%%%%%%%%%%%%%%%%%%%%%%%%%%%%%%%%%%%%%%%%%%
\subsection{Energy-momentum tensor of quarks and gluons}
\label{sect:enmomten}
%%%%%%%%%%%%%%%%%%%%%%%%%%%%%%%%%%%%%%%%%%%%%%%%%%%%%%%%%%%

The energy momentum tensor of the produced quarks and gluons has the form
\begin{equation}
T^{\mu\nu} = \int dP \, p^\mu p^\nu \left[ \sum_i \sum_f \left( G_{if}(x,p) + \bar{G}_{if}(x,p)\right) 
+\sum_{i,j=1}^3  \tilde{G}_{ij}(x,p)\right]
= \int dP \, p^\mu p^\nu S(\tau,w,p_\perp). \label{Tmunu1}
\end{equation}
Here we have introduced the function $S(\tau,w,p_\perp)$ to denote the expression in the square brackets in (\ref{Tmunu1}). The symmetry properties (\ref{symofg}) imply that $S(\tau,w,p_\perp)$ is an even function of $w$. Using this fact as well as Eqs. (\ref{binvv3})  and (\ref{binvm}) we find that the energy-momentum tensor (\ref{Tmunu1}) may be written in the form
\begin{eqnarray}
T^{\mu\nu} = (\varepsilon + P_\perp) U^\mu U^\nu - P_\perp g^{\mu\nu} + (P_\parallel-P_\perp) V^\mu V^\nu
\label{Tmunu2}
\end{eqnarray}
where
\begin{eqnarray}
\varepsilon(\tau) &=& \int dP \, \frac{v^2}{\tau^2}\,  S(\tau,w,p_\perp), \label{EPS} \\
P_\parallel(\tau) &=& \int dP \, \frac{w^2}{\tau^2}\,  S(\tau,w,p_\perp), \label{PL} \\
P_\perp(\tau) &=& \int dP \, \frac{p_\perp^2}{2}\,  S(\tau,w,p_\perp). \label{PT} 
\end{eqnarray}
The structure of the energy-momentum tensor (\ref{Tmunu2}) is typical for anisotropic systems, for example, see \cite{Florkowski:2008ag,Florkowski:2010cf,Martinez:2012tu}.

If we used the equilibrium distribution functions in (\ref{Tmunu1}) we would obtain the equilibrium energy-momentum tensor of the form
\begin{eqnarray}
T^{\mu\nu}_{\rm eq} = (\varepsilon_{\rm eq} + P_{\rm eq}) U^\mu U^\nu - P_{\rm eq} g^{\mu\nu}. 
\label{Tmunu2eq}
\end{eqnarray}
where
\begin{eqnarray}
\varepsilon_{\rm eq} = (6 N_f + 6) \frac{6T^4}{\pi^2}=\frac{36 (N_f+1) T^4}{\pi^2}, \quad P_{\rm eq} = \frac{1}{3} \varepsilon_{\rm eq}.
\label{BT4}
\end{eqnarray}
The factor $6 T^4/\pi^2$ describes the energy density of classical (Boltzmann) massless particles with the spin degeneracy 2. The Landau matching (\ref{LM}) is reduced in this case to the equation
\begin{eqnarray}
\varepsilon(\tau) = \varepsilon_{\rm eq}(\tau) = \frac{36 (N_f+1) T^4(\tau)}{\pi^2}.
\label{LM1}
\end{eqnarray}
Equation~(\ref{LM1}) allows us to determine the effective temperature of the system and to use it in the background distribution functions.

%%%%%%%%%%%%%%%%%%%%%%%%%%%%%%%%%%%%%%%%%%%%%%%%%%%%%%%%%%%
\subsection{Energy conservation for field and matter}
\label{sect:enfieldmatter}
%%%%%%%%%%%%%%%%%%%%%%%%%%%%%%%%%%%%%%%%%%%%%%%%%%%%%%%%%%%

The energy-momentum conservation law for the system of quarks, gluons and the chromoelectric field has the form
\begin{equation}
\partial _\mu T^{\mu \nu }(x)+\partial _\mu T_{\rm field}^{\mu \nu }(x)=0.
\label{enmomcon1}
\end{equation}
Here $T^{\mu\nu}$ is given by Eq.~(\ref{Tmunu2}), while $T_{\rm field}^{\mu \nu }$ is a diagonal energy-momentum tensor of the field
\begin{eqnarray}
T_{\rm field}^{\mu \nu }= 
 \left(
\begin{array}{cccc}
\varepsilon_{\rm field} & 0 & 0 & 0 \\
 0 & \varepsilon_{\rm field} & 0 & 0 \\
 0 & 0 & \varepsilon_{\rm field} & 0 \\
 0 & 0 & 0 & -\varepsilon_{\rm field}
\end{array} \right), \quad
\varepsilon_{\rm field} = \frac{1}{2}\, {\mbox{\boldmath $\cal E$}}^2.
\end{eqnarray}
One may notice that the field acts as matter whose transverse pressure is positive and equal to the field energy density $\varepsilon_{\rm field}$. On the other hand, the field longitudinal pressure is negative and equals $-\varepsilon_{\rm field}$. 

The total energy momentum tensor may be written in the form analogous to (\ref{Tmunu2}), namely
\begin{eqnarray}
T^{\mu\nu}_{\rm total} &=& (\varepsilon + P_\perp + 2 \varepsilon_{\rm field}) U^\mu U^\nu - (P_\perp+\varepsilon_{\rm field}) g^{\mu\nu} + (P_\parallel-P_\perp-2 \varepsilon_{\rm field}) V^\mu V^\nu \nonumber \\
&\equiv& (\varepsilon_{\rm total} + P^\perp_{\rm total}) \, U^\mu U^\nu - P^\perp_{\rm total} g^{\mu\nu} + (P^\parallel_{\rm total}-P^\perp_{\rm total}) V^\mu V^\nu.
\label{Tmunu3}
\end{eqnarray}
The energy-momentum conservation (\ref{enmomcon1}) implies that the following equation should be always satisfied 
\begin{eqnarray}
\frac{d\varepsilon_{\rm field}}{d\tau}+
\frac{d\varepsilon}{d\tau}=
- \frac{\varepsilon+P_\parallel}{\tau}.
\label{depsdtau}
\end{eqnarray}
We note that in the case without the field, the first term on the left-hand-side of Eq.~(\ref{depsdtau}) is absent and (\ref{depsdtau}) is reduced to the equation used in anisotropic hydrodynamics for one-dimensional and boost-invariant expansion.

%%%%%%%%%%%%%%%%%%%%%%%%%%%%%%%%%%%%%%%%%%%%%%%%%%%%%%%%%%%
\subsection{Equations of motion}
\label{sect:EOM}
%%%%%%%%%%%%%%%%%%%%%%%%%%%%%%%%%%%%%%%%%%%%%%%%%%%%%%%%%%%

In this Section we show that the field equations (\ref{Maxwell})--(\ref{convcur}) guarantee that the energy and momentum of our system are conserved quantities. The $\nu =0$ component of the energy-momentum conservation equation includes the terms 
\begin{equation}
\partial _\mu T_{\rm field}^{\mu 0}=\frac \partial {\partial t}\left( \frac 12{%
\mbox{\boldmath $\cal E$}}^2\right) ={\mbox{\boldmath $\cal E$}\cdot }\frac{%
\partial {\mbox{\boldmath $\cal E$}}}{\partial t}=-{\bf F}_{}^{30}\cdot 
\frac{\partial {\bf F}_{}^{03}}{\partial t}  \label{enmomcon3a}
\end{equation}
and 
\begin{eqnarray}
\partial _\mu T_{}^{\mu 0} &=&  \int dPp^0p^\mu \partial _\mu \left[
\sum_i\sum_f\left( G_{if}+\bar{G}_{if}\right) +\sum_{i,j=1}^3%
\widetilde{G}_{ij}\right]  \nonumber \\
&=&-g{\bf F}_{}^{\mu \nu }\cdot \int dP\ p^0\ p_\nu \ \partial _\mu ^p\left[
\sum_i{\mbox{\boldmath $\epsilon$}}_i\sum_f\left( G_{if}-\bar{G}%
_{if}\right)  
+ \sum_{i,j=1}^3{\mbox{\boldmath $\eta$}}_{ij}\widetilde{G}_{ij}%
\right] \nonumber \\
&&
+2\int dP\ p^0\left[ \sum_i\sum_f\frac{dN_{if}}{d\Gamma_{\rm inv} }%
+\sum_{i>j}\frac{d\widetilde{N}_{ij}}{d\Gamma_{\rm inv} }\right] . \nonumber \\
\label{enmomcon3b}
\end{eqnarray}
In Eq.~(\ref{enmomcon3b}) we used the kinetic equations (\ref{kineq})--(\ref{kineg}) and applied the Landau matching (\ref{LM}). This allows us to write the $\nu =0$ component of the total energy-momentum conservation law in the form
\begin{eqnarray}
&& {\bf F}_{}^{30}\cdot \frac{\partial {\bf F}_{}^{03}}{\partial t} = {\bf F}_{}^{30}\cdot g\int dPp^3\left[ \sum_i{\mbox{\boldmath $\epsilon$}}%
_i\sum_f\left( G_{if}-\bar{G}_{if}\right) 
 +\sum_{i,j=1}^3{%
\mbox{\boldmath $\eta$}}_{ij}\widetilde{G}_{ij}\right]  
\nonumber \\
&& + \, {\bf F}_{}^{30}\cdot \int dP\ \left[ \sum_i\frac{2p^0{\mbox{\boldmath $\epsilon$}}_i}{\ {\mbox{\boldmath $\epsilon$}}_i\cdot {\bf F%
}_{}^{30}}\sum_f\frac{dN_{if}}{d\Gamma_{\rm inv} }
+\sum_{i>j}\frac{2p^0{%
\mbox{\boldmath $\eta$}}_{ij}^{}}{{\mbox{\boldmath
$\eta$}}_{ij}\cdot {\bf F}_{}^{30}}\frac{d\widetilde{N}_{ij}}{d\Gamma_{\rm inv} }%
\right] .  \nonumber \\
\label{enmomcon3c}
\end{eqnarray}
We thus see, that the field equations (\ref{Maxwell})--(\ref{convcur}) with the definitions of the dipole moments (\ref{dipols}) represent a sufficient condition to have the total energy of the system conserved (note that the delta functions present in the production rates generate zero longitudinal momenta at $z=0$). Similarly, we may analyze the $\nu =3$ component of the energy-momentum conservation law and obtain the same conclusion.

%%%%%%%%%%%%%%%%%%%%%%%%%%%%%%%%%%%%%%%%%%%%%%%%%%%%%%%%%%%
\section{Results}
\label{sect:results}
%%%%%%%%%%%%%%%%%%%%%%%%%%%%%%%%%%%%%%%%%%%%%%%%%%%%%%%%%%%

\subsubsection{Initial conditions and model parameters}

In this Section we present our main results. In the numerical calculations we use two values of the parameter $k$ ($k=5$ and $k=10$) and three values of the parameter $4 \pi{\bar \eta}$, see Eqs.~(\ref{taueq}) and (\ref{etabars}). The extra cases defined by the condition $4\pi{\bar \eta}=\infty$ correspond to the situations where the collision terms are absent and the system's dynamics is determined by the mean field only (Boltzmann-Vlasov limit). For sake of simplicity, the masses of quarks are neglected and we take into account only two quark flavors, $N_f=2$ (the incorporation of the finite quark masses into the present formalism is quite straightforward but all the equations become much more intricate, see, for example, Ref.~\cite{Dyrek:1988eb}). In addition, we assume that the initial field is spanned by the gluons with the charge ${\mbox{\boldmath $\eta$}}_{12}$,
\begin{equation}
{\mbox{\boldmath $\cal E$}}_0 = \left({\cal E}^0_{(3)},{\cal E}^0_{(8)}=0\right) =
\sqrt{\frac{6\sigma _q}{\pi r^2}}k{\mbox{\boldmath $\eta$}}_{12} . \label{inite0}
\end{equation}
In this case, the second component of the field is always zero. With the initial condition (\ref{inite0}) we solve Eq.~(\ref{biMaxwell}) and obtain the functions ${\bf h}(\tau)=(h_{(3)}(\tau),h_{(8)}(\tau)=0)$ and $T(\tau)$. The knowledge of $h_{(3)}(\tau)$ and $T(\tau)$ allows us to calculate all other interesting physical quantities.

\begin{figure}[t]
\begin{center}
\subfigure{\includegraphics[angle=0,width=0.49\textwidth]{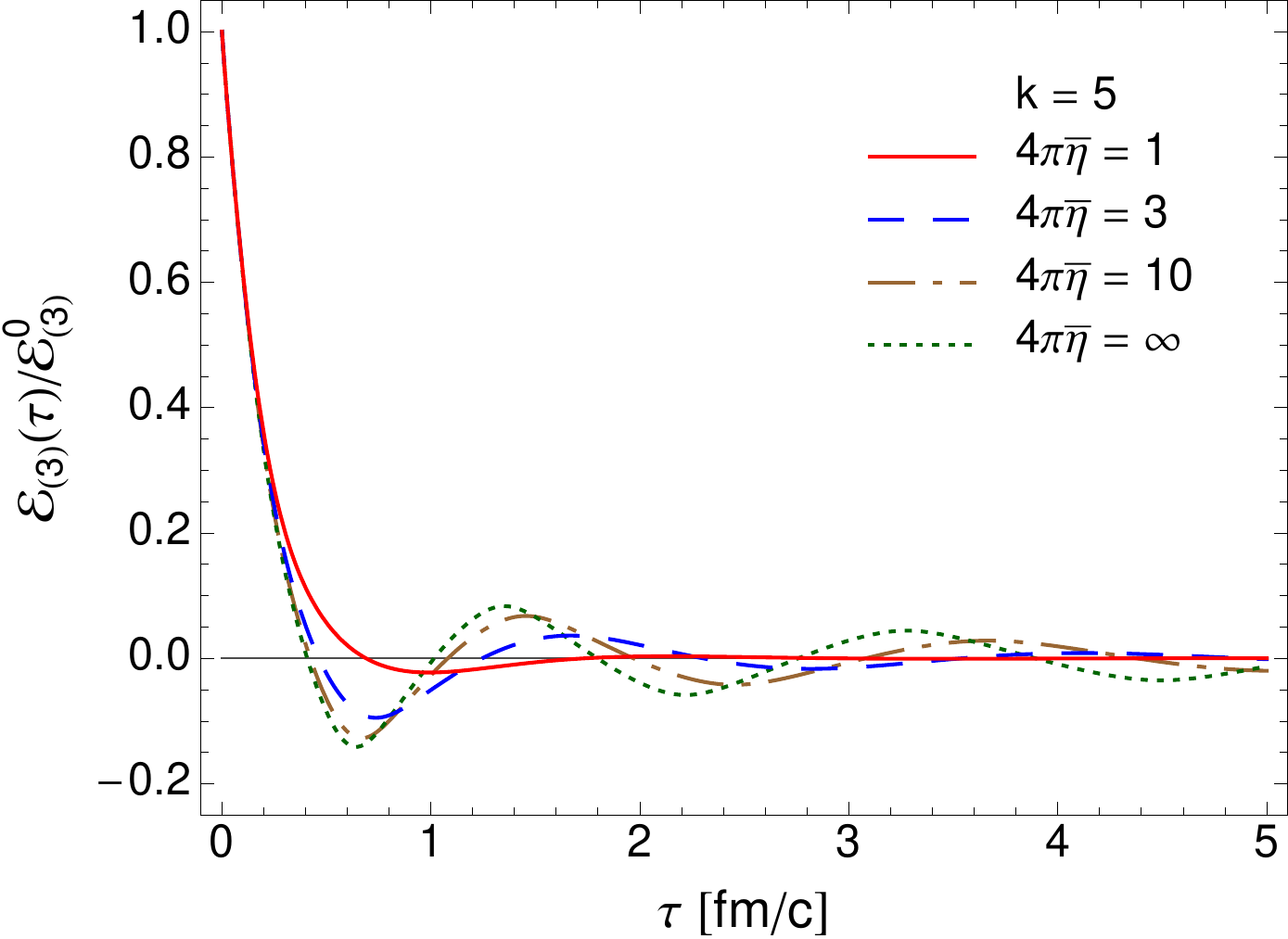}} 
\subfigure{\includegraphics[angle=0,width=0.49\textwidth]{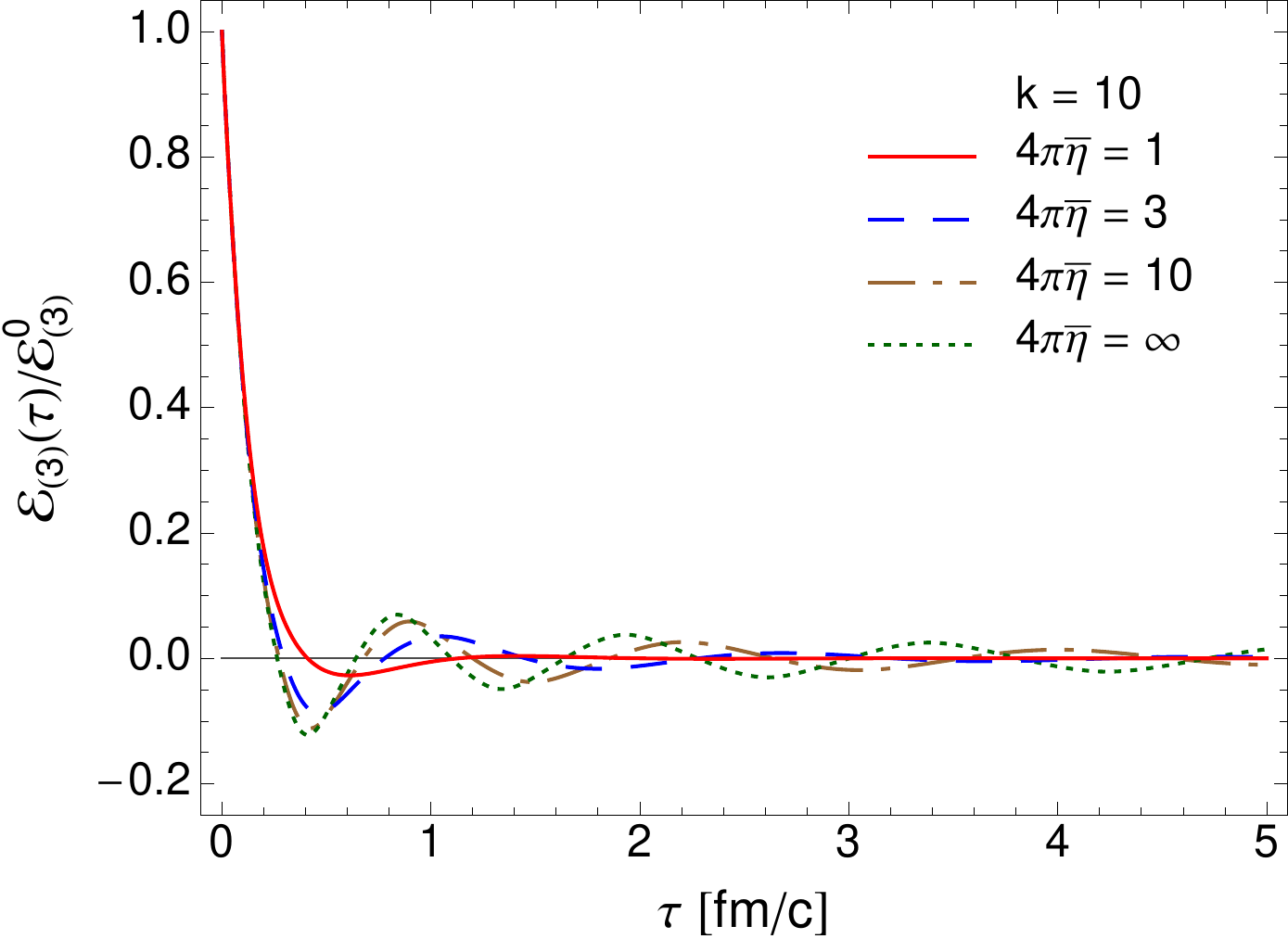}}
\end{center}
\caption{(Color online) Time dependence of the chromoelectric field for different values of the viscosity in the case $k=5$ (left) and $k=10$ (right). The values of the field are normalized to its initial value.
}
\label{fig:EF}
\end{figure}

\subsubsection{Oscillations vs. damping of the color fields}

In Fig.~\ref{fig:EF} we show the time dependence of the chromoelectric field normalized to its initial value, for $k=5$ (left) and $k=10$ (right). Different curves describe our results obtained for different values of the viscosity. In the case $4\pi{\bar \eta}=\infty$ and $k=5$ (dotted green line in the left part of Fig.~\ref{fig:EF}) we reproduce the result obtained earlier in Ref.~\cite{Bialas:1987en}. This result describes oscillations of the chromoelectric field which are slowly damped due to the longitudinal expansion of the system. Similarly, in the case $4\pi{\bar \eta}=\infty$ and $k=10$ (dotted green line in the right part of Fig.~\ref{fig:EF}) we deal again with the oscillations of the chromoelectric field --- they are faster than those found in the case $4\pi{\bar \eta}=\infty$ and $k=5$. The increase of the field frequency with the increasing strength of the initial chromoelectric field was observed earlier for $k \leq 5$ in Ref.~\cite{Bialas:1987en}. 

A new aspect of the present work is the inclusion of the viscosity effects which are characterized by the parameter $4\pi{\bar \eta}$. As the viscosity of the system decreases, the collisions between particles become more frequent, the system becomes more dissipative, and the oscillations of the chromoelectric field are more and more damped. For $4\pi{\bar \eta}=1$ the oscillations practically disappear for both $k=5$ and $k=10$ (solid red lines in the left and right parts of Fig.~\ref{fig:EF}). Nevertheless, for larger values of the viscosity (for example, for $4\pi{\bar \eta}=3$) the effects of collisions do not seem to be efficient enough to completely damp down the plasma oscillations. Only if the viscosity is defined by the KSS bound, the oscillations may be neglected.

It is interesting to note that the dependence of the relaxation time on the inverse of the effective temperature makes $\tau_{\rm eq}$ on average smaller in the case $k=10$ than in the case $k=5$. This leads to similarities between the cases $k=5$ and $k=10$ because larger initial fields in the case $k=10$ are damped faster by the collisions, while the smaller initial fields in the case $k=5$ are damped slower.

\begin{figure}[t]
\begin{center}
\subfigure{\includegraphics[angle=0,width=0.49\textwidth]{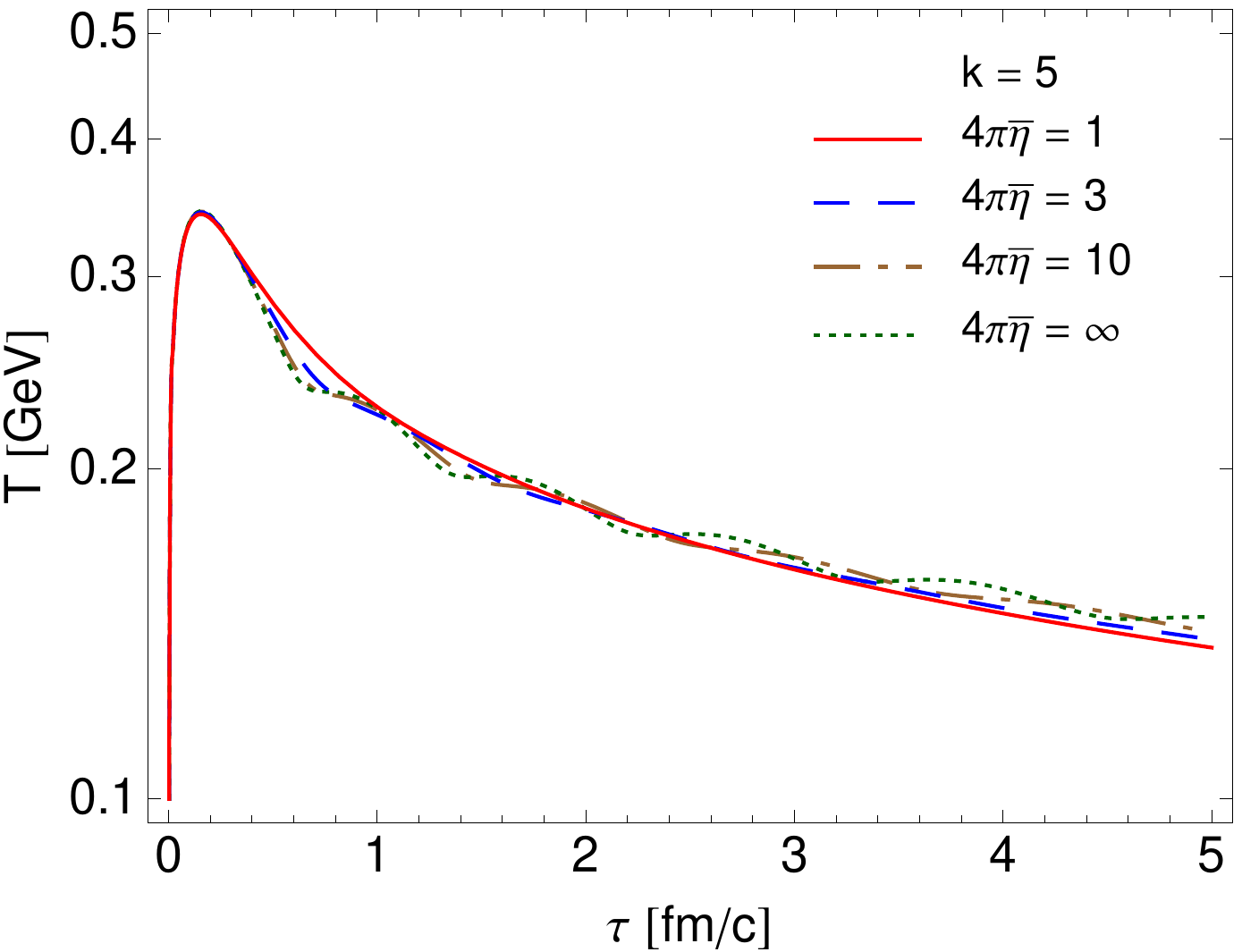}} 
\subfigure{\includegraphics[angle=0,width=0.49\textwidth]{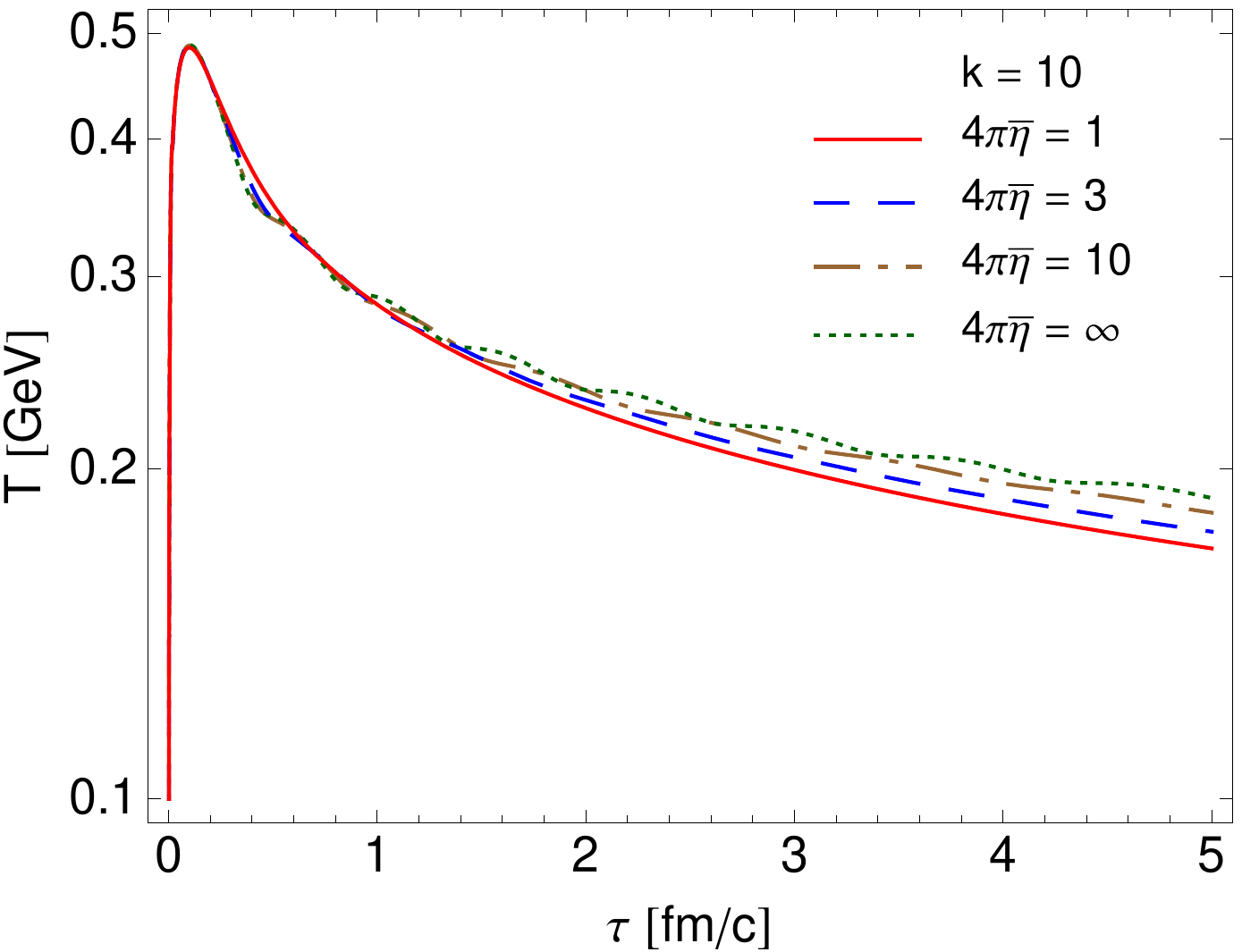}}
\end{center}
\caption{(Color online) Time dependence of the effective temperature $T$ for different values of the viscosity in the case $k=5$ (left) and $k=10$ (right).
}
\label{fig:T}
\end{figure}

\begin{figure}[t]
\begin{center}
\subfigure{\includegraphics[angle=0,width=0.49\textwidth]{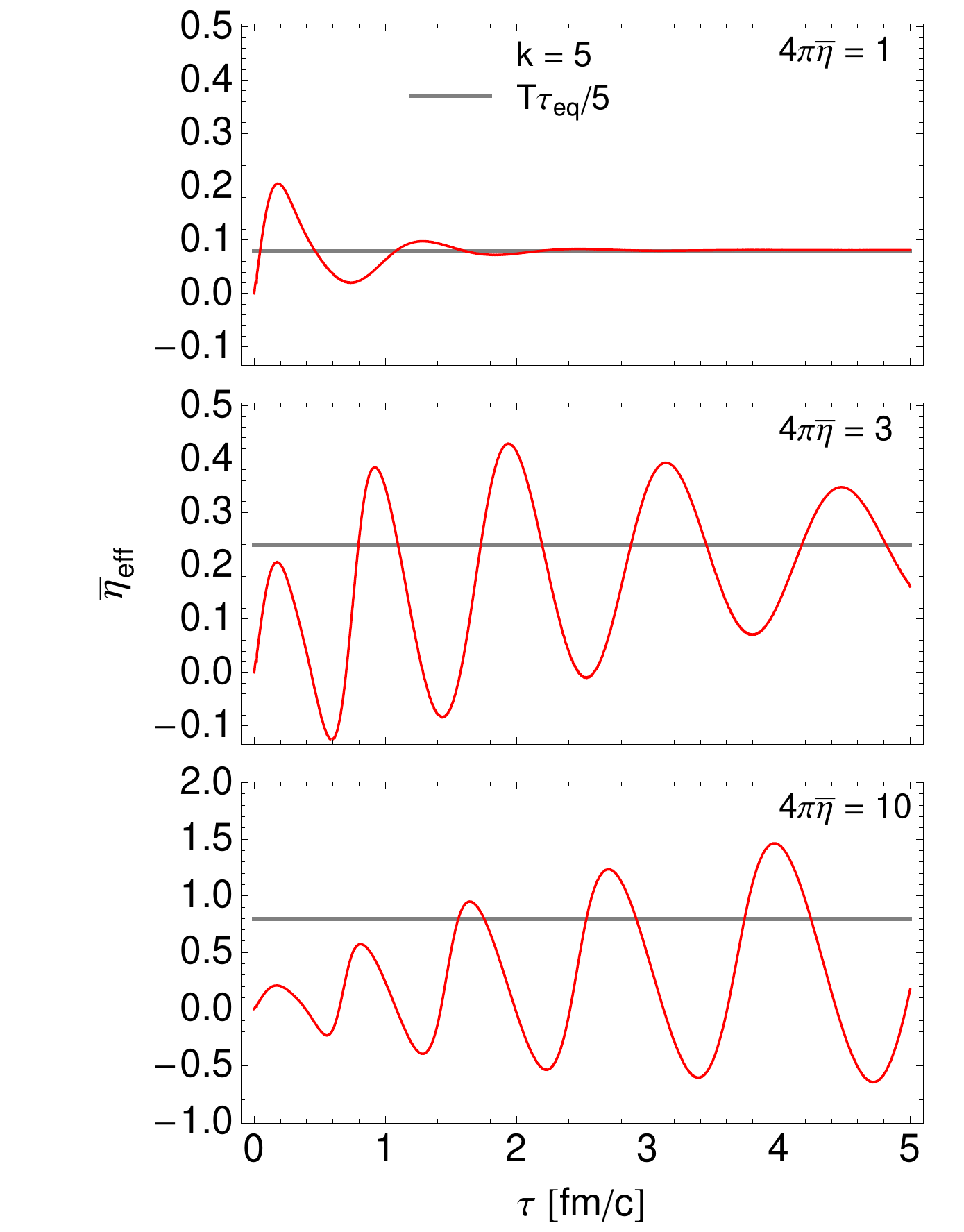}} 
\subfigure{\includegraphics[angle=0,width=0.49\textwidth]{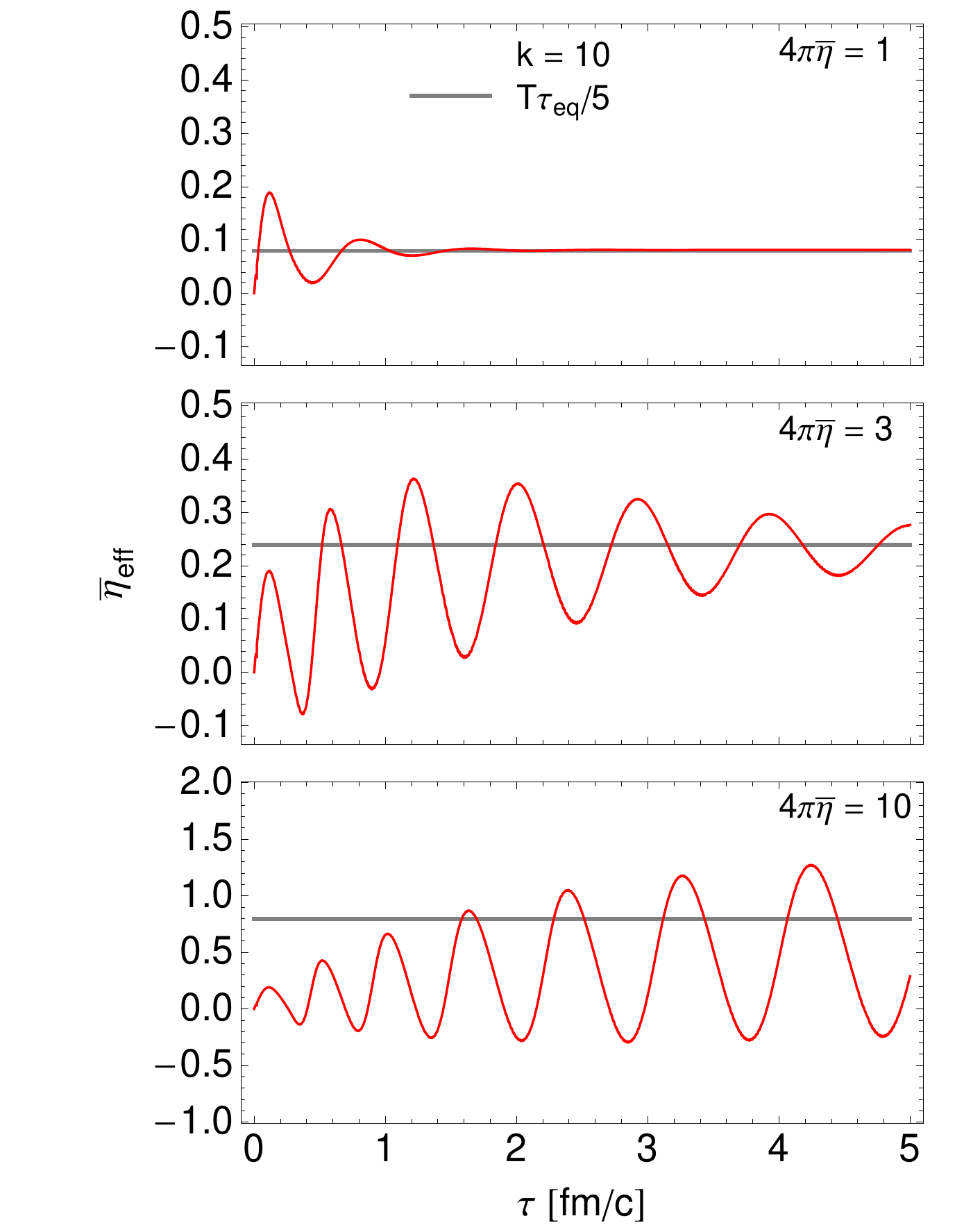}}
\end{center}
\caption{(Color online) Time dependence of the effective viscosity ${\bar \eta}_{\rm eff}$ for different values of the parameter $4\pi{\bar \eta}$ in the case $k=5$ (left) and $k=10$ (right).
}
\label{fig:etaeff}
\end{figure}

\subsubsection{Effective temperature and effective viscosity}

Figure~\ref{fig:T} shows the time dependence of the effective temperature $T(\tau)$ of the system for $k=5$ (left) and $k=10$ (right). Again, different curves describe our results obtained for different values of the viscosity. The effective temperature is a measure of the energy density of the produced quarks and gluons, see Eq.~(\ref{LM1}).  Initially, it is equal to zero, however, due to the fast decay of the initial chromoelectric field, it grows rapidly at the very early stages of the evolution of the system.  After the initial rapid growth, the effective temperature decreases in a qualitatively similar way to that predicted in the Bjorken model \cite{Bjorken:1982qr}; one may check that  $T(\tau) \approx \tau^{-1/3}$. However, the presence of both viscosity and color fields leads to noticeable deviations from the Bjorken-scaling behavior. 

In order to quantitatively characterize the system's behavior at later times we assume that the plasma may be characterized by the first-order viscous-hydrodynamics equations. In the case of one-dimensional boost-invariant expansion, the ratio of the shear viscosity to entropy density is  connected with the system's effective temperature $T(\tau)$ and its time derivative $dT(\tau)/d\tau$ through the formula
\begin{eqnarray}
\frac{dT}{d\tau}+\frac{T}{3\tau}=\frac{4 {\bar \eta}_{\rm eff}}{9 \tau^2},
\label{eta-fo}
\end{eqnarray}
We use the numerical results for the functions $T(\tau)$ shown in Fig.~\ref{fig:T} and substitute them into the left-hand-side of Eq.~(\ref{eta-fo}) in order to calculate the effective viscosity ${\bar \eta}_{\rm eff}$. The functions ${\bar \eta}_{\rm eff}(\tau)$ obtained for different values of the parameters $k$ and $4\pi{\bar \eta}$ are shown in Fig.~\ref{fig:etaeff}. 

At first, in the two upper parts of Fig.~\ref{fig:etaeff} one can notice that in the minimum viscosity case (defined by the condition $4\pi{\bar \eta}=1$) the effective viscosity of the system starts to agree very well with the viscosity parameter ${\bar \eta}$ after 1--2 fm/c, for both $k=5$ and $k=10$. In practice, this means that for $\tau >$ 1~fm/c our complicated system of fields and particles is very well described by the first-order viscous hydrodynamics (and, consequently, also by the second order hydrodynamics). 

On the other hand, for larger values of the viscosity, for example, in the cases $4\pi{\bar \eta}=3$ and $4\pi{\bar \eta}=10$ (see the two middle and two lower parts of Fig.~\ref{fig:etaeff}) the effective viscosity ${\bar \eta}_{\rm eff}$ differs from the value of ${\bar \eta}$. In these cases the collisions in the plasma become inefficient to damp down the plasma oscillations. The presence of such oscillations brings in differences between the kinetic and viscous-hydrodynamics descriptions, and indicates that the viscous-hydrodynamics description after 1--2 fm/c is not completely satisfactory if $4\pi{\bar \eta} \geq 3$.~\footnote{The same holds if one uses the second order hydrodynamics. The oscillations of the plasma parameters imply strong variations in the initial conditions for hydrodynamics and make the pure hydrodynamic predictions very unstable.} The presence of the color fields at this stage suggests that the viscous hydrodynamics should be extended to include transport phenomena connected with color conductivity.

However, it is interesting to observe in the case $4\pi{\bar \eta}=3$  that
although ${\bar \eta}_{\rm eff}$ and ${\bar \eta}$ are different, after 1--2 fm/c the effective viscosity starts to fluctuate around the constant value corresponding to ${\bar \eta}$. In this case, one may consider averaging over different color-flux-tubes which washes out the oscillations in such a way that the averaged system may be effectively well described by the viscous hydrodynamics.  

\subsubsection{Energy density, longitudinal pressure, transverse pressure}

\begin{figure}[t]
\begin{center}
\subfigure{\includegraphics[angle=0,width=0.49\textwidth]{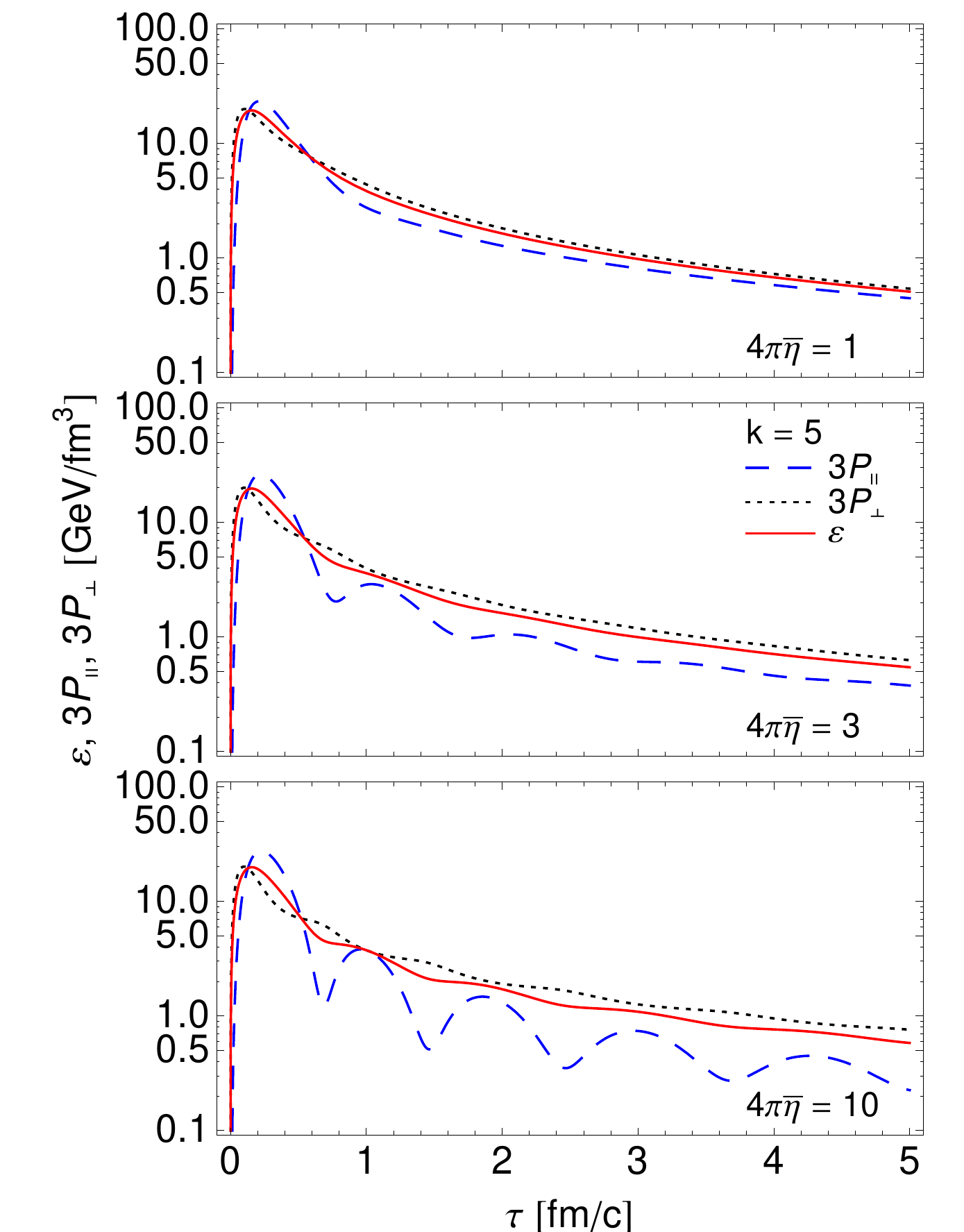}} 
\subfigure{\includegraphics[angle=0,width=0.49\textwidth]{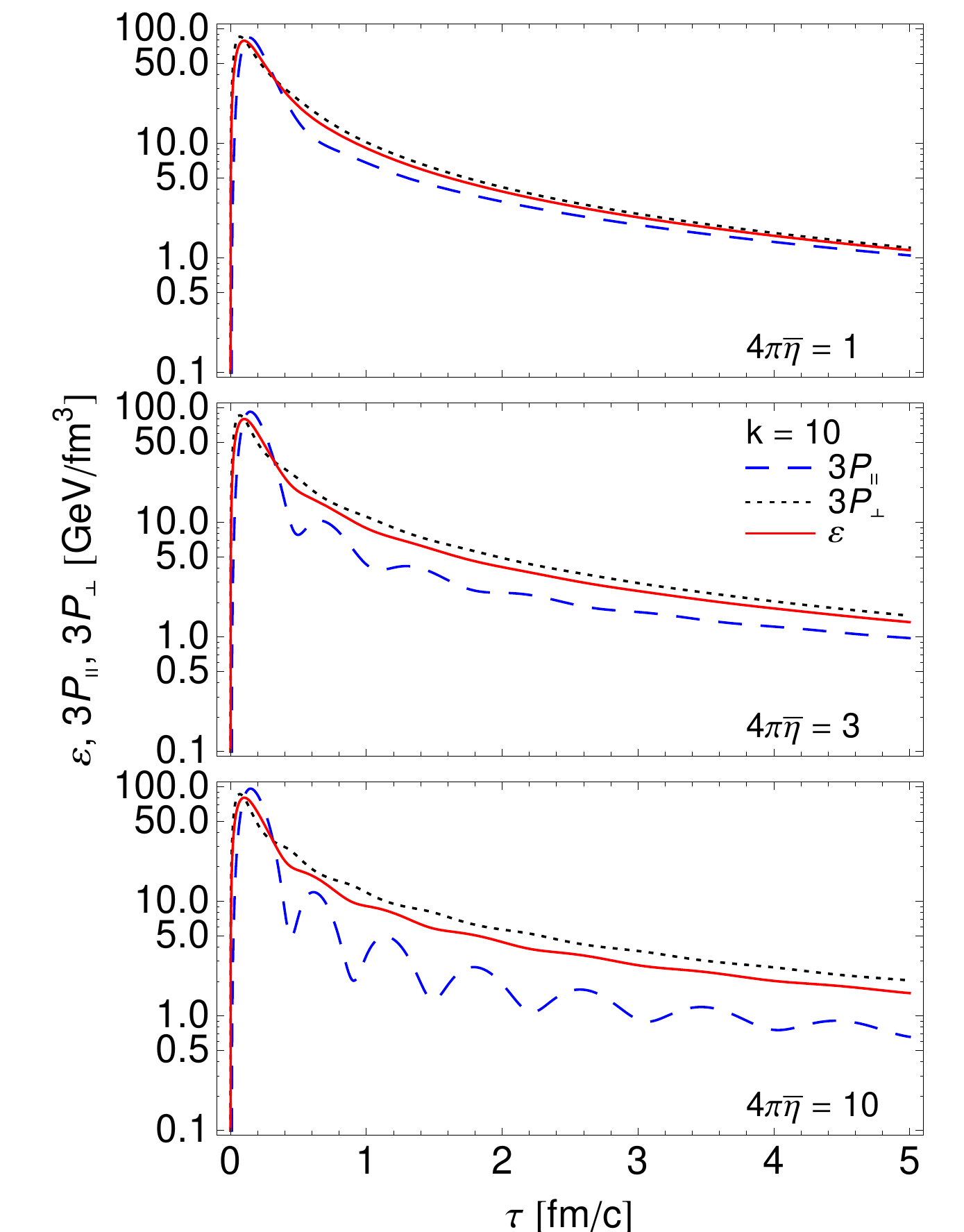}} 
\end{center}
\caption{(Color online) Time dependence of the energy density, longitudinal pressure, and transverse pressure of the produced quark-gluon plasma again in the case $k=5$ (left) and $k=10$ (right).
}
\label{fig:EP}
\end{figure}

In Fig.~\ref{fig:EP} we show the time dependence of the energy density, $\varepsilon(\tau)$, the longitudinal pressure, $P_\parallel(\tau)$, and the transverse pressure, $P_\perp(\tau)$, of the produced quarks and gluons. These quantities have been calculated according to the formulas given in the Appendix \ref{app:energy_density}, see discussions following Eqs.~(\ref{eps1}), (\ref{PL1}), and (\ref{PT1}). Here we use again the values $k=5$ (left) and $k=10$ (right). In order to have similar time asymptotic behavior for all thermodynamics-like quantities shown in the figure, the two pressures are multiplied by the factor of 3. Three panels in the left and right parts of Fig.~\ref{fig:EP}  correspond to three different values of the parameter $4\pi{\bar \eta}$. 

The two upper parts of Fig.~\ref{fig:EP} may be interpreted as an illustration of the fast and almost complete thermalization of matter --- the energy density of the plasma as well as the two rescaled pressures approach very fast each other. This result indicates that the viscosity corresponding to the choice $4\pi{\bar \eta}=1$ is efficient to equilibrate the system within 1--2~fm/c. Nevertheless, small differences between the energy density and the two rescaled pressures remain. They are caused by the non-zero viscosity included in the kinetic approach, which has been discussed above.

The two middle panels of Fig.~\ref{fig:EP} show our results obtained for $4\pi{\bar \eta}=3$. In this case the differences between the functions $\varepsilon(\tau)$, $3P_\parallel(\tau)$, and $3P_\perp(\tau)$ are larger. In particular, the longitudinal pressure (dashed lines) is clearly below the transverse pressure (dotted lines) for $\tau >$ 1 fm/c. This is certainly an effect of the non-zero viscosity of the system which tends to lower the longitudinal pressure and to increase the transverse pressure. 

The differences between the transverse and longitudinal pressures become even larger in the case $4\pi{\bar \eta}=10$ which is shown in the two lower panels of Fig.~\ref{fig:EP}. In this case, the longitudinal pressure is not only much smaller than the transverse pressure but it strongly oscillates in time. The survival of such oscillations indicates that the collision rate is not sufficiently fast to destroy collective phenomena in the plasma.

\subsubsection{Longitudinal vs. transverse pressure}

\begin{figure}[t]
\begin{center}
\subfigure{\includegraphics[angle=0,width=0.49\textwidth]{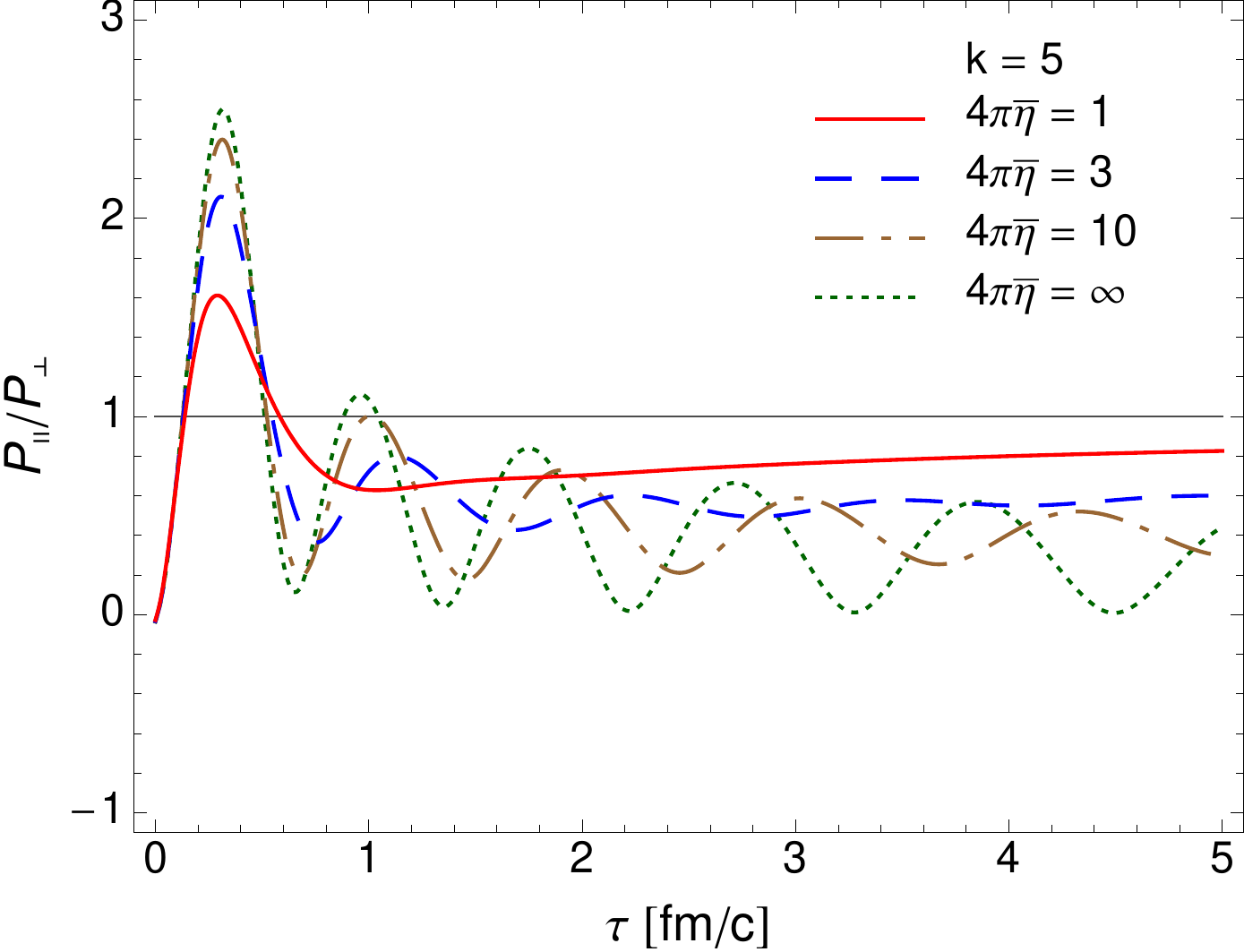}} 
\subfigure{\includegraphics[angle=0,width=0.49\textwidth]{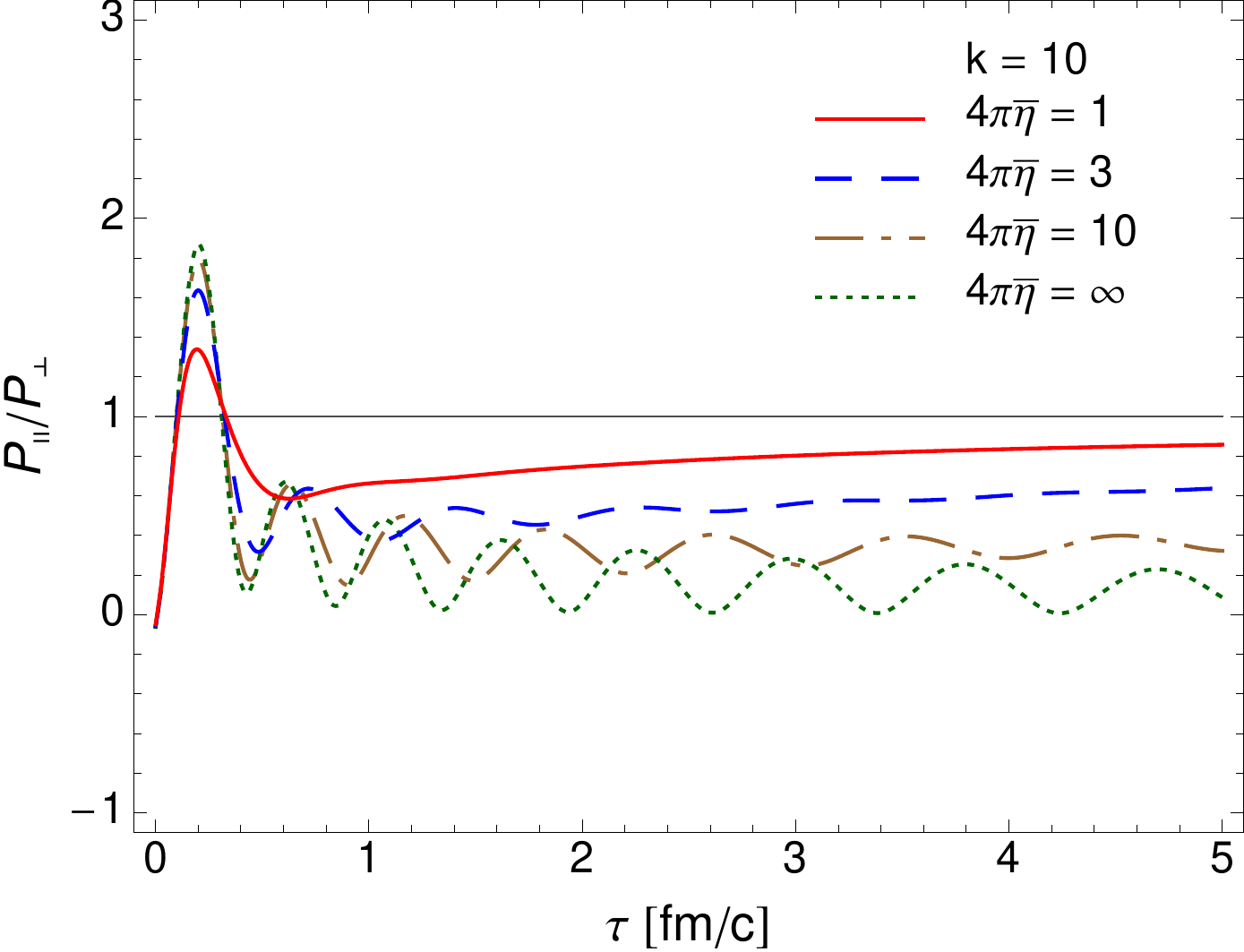}} 
\end{center}
\caption{(Color online) Time dependence of the ratio $P_\parallel(\tau)/P_\perp(\tau)$ for $k=5$ (left) and $k=10$, and for different values of the viscosity.
}
\label{fig:PLoverPT}
\end{figure}

In order to follow in more detail the system's approach towards local thermodynamic equilibrium, in Fig.~\ref{fig:PLoverPT} we show the time dependence of the ratio $P_\parallel(\tau)/P_\perp(\tau)$. At $\tau=0$ the longitudinal pressure is zero. This behavior is connected with the fact that tunneling particles emerge from the vacuum with vanishing longitudinal momenta (see the Dirac delta functions $\delta(p_\parallel)$ or $\delta(w)$ appearing in the quark and gluon production rates, Eqs.~(\ref{qrate}), (\ref{grate}) and (\ref{binvterms})). In the collisionless case, $4\pi{\bar \eta}=\infty$, the ratio of the two pressures strongly oscillates and its average value is significantly smaller than 1. With decreasing viscosity, the ratio $P_\parallel(\tau)/P_\perp(\tau)$ gets closer to unity.

\begin{figure}[t]
\begin{center}
\subfigure{\includegraphics[angle=0,width=0.49\textwidth]{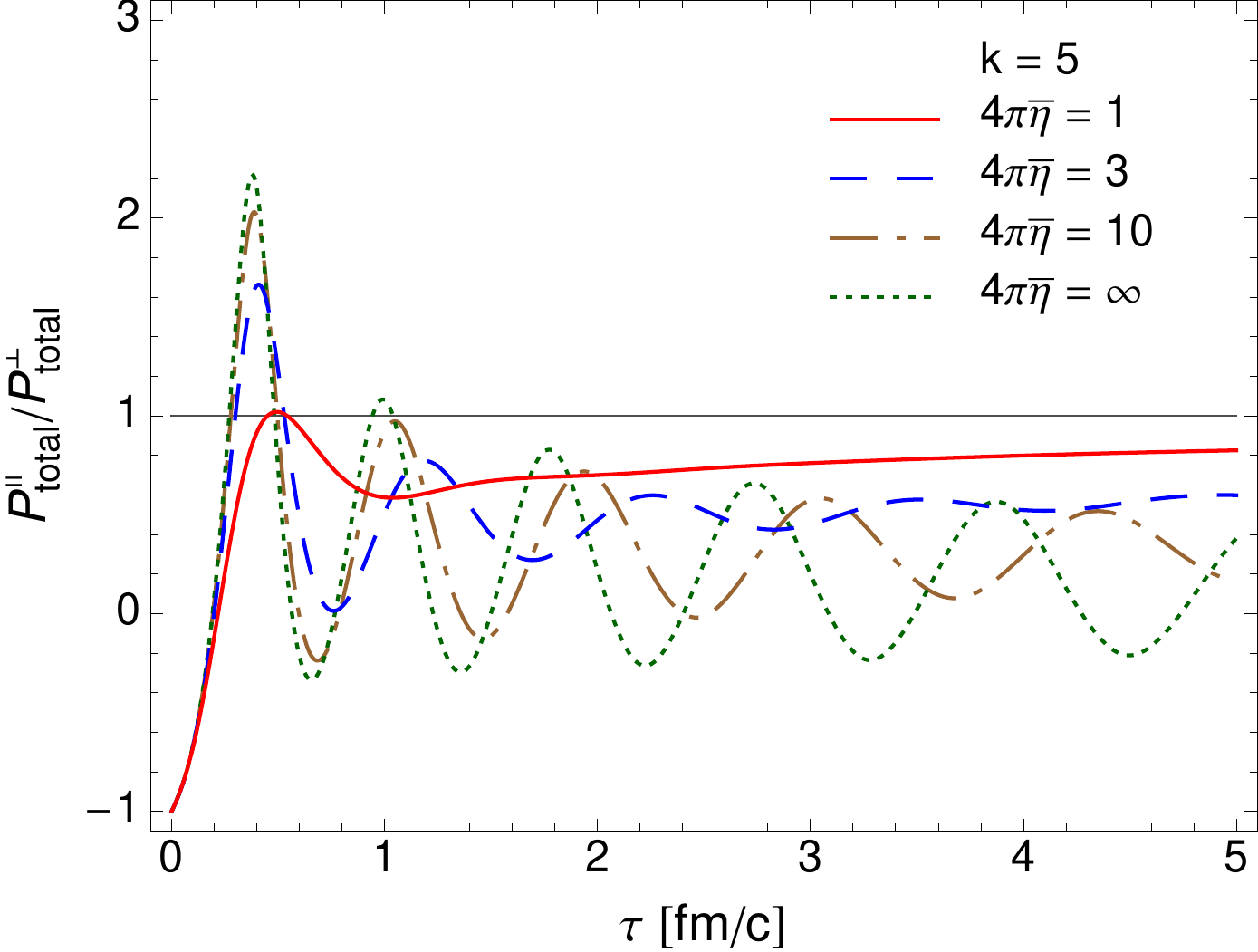}} 
\subfigure{\includegraphics[angle=0,width=0.49\textwidth]{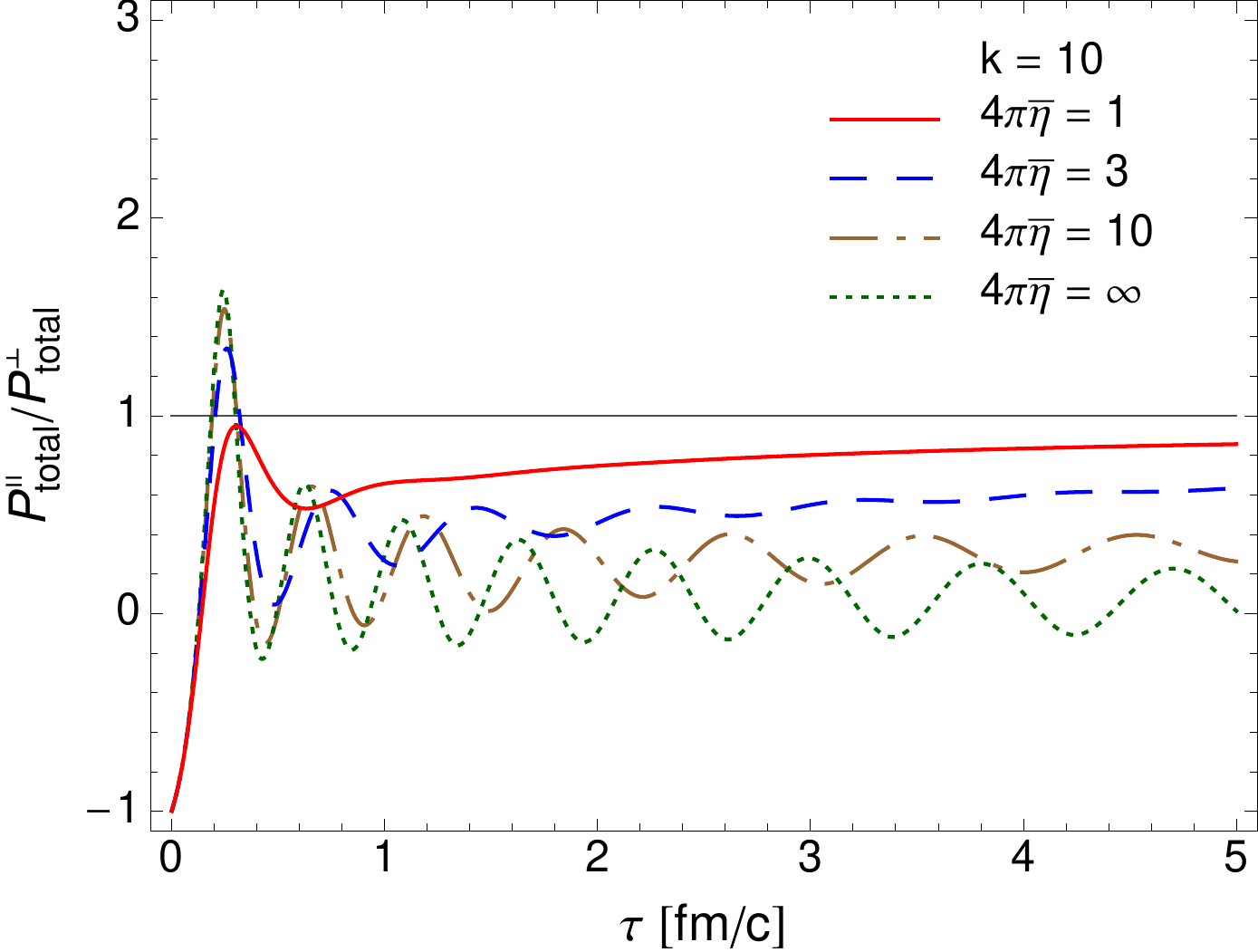}} 
\end{center}
\caption{(Color online) Same as Fig.~\ref{fig:PLoverPT} but the pressures include the contributions from the field.
}
\label{fig:PLoverPT2}
\end{figure}

In Fig.~\ref{fig:PLoverPT} the two pressures characterize the produced quarks and gluons. It is interesting to note, however, that the chromoelectric field gives extra contributions to the longitudinal and transverse pressures, and the complete expressions for the pressures should include the field parts. The ratio of the total (particles+field) longitudinal pressure and the total (particles+field) transverse pressure is shown in Fig.~\ref{fig:PLoverPT2}. Since the longitudinal chromoelectric field gives a negative contribution to the longitudinal pressure and a positive contribution to the transverse pressure, the main effect of the field is to lower the $P_\parallel(\tau)/P_\perp(\tau)$ ratio calculated for the system of particles only. This effect is, of course, the strongest at the initial stages when the color fields are the strongest. For later times, when the field decays and becomes negligible, the field contributions to the energy-momentum tensor become irrelevant and the results shown in Figs.~\ref{fig:PLoverPT} and \ref{fig:PLoverPT2} are very much similar.

%%%%%%%%%%%%%%%%%%%%%%%%%%%%%%%%%%%%%%%%%%%%%%%%%%%%%%%%%%%
\section{Conclusions}
\label{sect:concl}
%%%%%%%%%%%%%%%%%%%%%%%%%%%%%%%%%%%%%%%%%%%%%%%%%%%%%%%%%%%

In this paper we have analyzed equilibration of the quark-gluon plasma produced by decays of color flux tubes possibly created at the early stages of relativistic heavy-ion collisions. A novel feature of our approach is the implementation of the viscosity of the produced quark-gluon plasma in terms of a constant ratio of the shear viscosity coefficient to the entropy density, ${\bar \eta} = \eta/\sigma$ = const. For constant ${\bar \eta}$, the relaxation time in the collision terms becomes a function of the effective temperature of the plasma, and the numerical analysis of the kinetic equations becomes much more involved than that applied in the case where the relaxation time is a constant.

In our numerical calculations we have used realistic values of the initial field strength and the viscosity. The initial field strength is chosen in such a way that the effective temperature of the produced plasma reaches values expected at RHIC and the LHC, $T_{\rm max} \sim 300-500$ MeV.  For the lowest (KSS) value of the ratio of the shear viscosity to the entropy density, $4\pi{\bar \eta} = 1$, the analyzed system approaches the viscous-hydrodynamics regime within  1--2 fm/c. On the other hand, for larger values of the viscosity, $4\pi{\bar \eta} \geq 3$, the collisions in the plasma are not efficient to destroy collective phenomena in the plasma, which manifest themselves as oscillations of different plasma parameters. The presence of such oscillations after the first 1--2 fm/c brings in differences between the kinetic and viscous hydrodynamic descriptions, which suggests that the viscous-hydrodynamics description is not complete at this stage if $4\pi{\bar \eta} \geq 3$ and should be extended to include dissipative phenomena connected with color conductivity.

\begin{acknowledgments}

R.R. was supported in part by the Polish National Science Center grant with decision No. DEC-2012/07/D/ST2/02125 and the Foundation for Polish Science. W.F. was supported in part by the Polish National Science Center grant with decision no. DEC-2012/06/A/ST2/00390.                    

\end{acknowledgments}

\appendix

%%%%%%%%%%%%%%%%%%%%%%%%%%%%%%%%%%%%%%%%%%%%%%%%%%%%%
\section{Energy density, longitudinal pressure and transverse pressure}
\label{app:energy_density}
%%%%%%%%%%%%%%%%%%%%%%%%%%%%%%%%%%%%%%%%%%%%%%%%%%%%%

In this Section we give details of our calculations of the energy density and the two pressures, which have been presented in Sec.~\ref{sect:results}.

%%%%%%%%%%%%%%%%%%%%%%%%%%%%%%%%%%%%%%%%%%%%%%%%%%%%%
\subsection{Energy density}
%%%%%%%%%%%%%%%%%%%%%%%%%%%%%%%%%%%%%%%%%%%%%%%%%%%%%

Our starting point for the calculation of the energy density of quarks and gluons is Eq.~(\ref{EPS}),
\begin{eqnarray}
\varepsilon(\tau) &=& \int dP \, \frac{v^2}{\tau^2} \left[ N_f \sum_{i} \left( G_{i}(x,p) + \bar{G}_{i}(x,p)\right) +\sum_{i>j} \left(
 \tilde{G}_{ij}(x,p) +  \tilde{G}_{ji}(x,p) \right)
 \right] \nonumber \\
 &=& \frac{2}{\tau^2} \int \, dw\,  \int d^2p_\perp \,v \left[ N_f \sum_{i}  G_{i}(x,p)  + \sum_{i>j}  \tilde{G}_{ij}(x,p)  \right]  .
  \label{eps1}
\end{eqnarray}
Since in the numerical calculations we take into account only two massless flavors, we have changed here our notation from $G_{if}(x,p)$ to $G_{i}(x,p)$, and replaced the sum by the factor $N_f$. We have also used the symmetry properties of the distribution functions with respect to the change $w \rightarrow -w$, see Eq.~(\ref{symofg}). In the next step we use the general forms of the solutions of the kinetic equations (\ref{formsol}), and substitute them into Eq.~(\ref{eps1}). In this way we obtain
\begin{eqnarray}
\varepsilon(\tau) &=& \frac{N_f}{4\pi^3 \tau^2} \, 
\int_0^\tau d\tau^\prime D(\tau,\tau^\prime)
\sum_{i} 
\left[2 \tau^\prime \Lambda_i^2(\tau^\prime)
D^-(\beta_i(\tau,\tau^\prime)) \sqrt{\Delta h_i^2(\tau,\tau^\prime)+\frac{\Lambda_i(\tau^\prime) \tau^2}{\pi}}
+ \Sigma^\varepsilon_i(\tau,\tau^\prime) \right]
\nonumber \\
&& +\frac{1}{4\pi^3 \tau^2} \, 
\int_0^\tau d\tau^\prime D(\tau,\tau^\prime)
\sum_{i>j} 
\left[2 \tau^\prime \Lambda_{ij}^2(\tau^\prime)
D^+(\beta_{ij}(\tau,\tau^\prime)) \sqrt{\Delta h_{ij}^2(\tau,\tau^\prime)+\frac{\Lambda_{ij}(\tau^\prime) \tau^2}{\pi}}
+ \Sigma^\varepsilon_{ij}(\tau,\tau^\prime) \right]. \label{eps2}
\end{eqnarray}
The function $\beta(\tau,\tau^\prime)$ has been already defined by Eq.~(\ref{beta}), while the functions $D^{\pm}(\beta)$ have been defined by Eq.~(\ref{Dpm}). The function $\Sigma^\varepsilon(\tau,\tau^\prime)$  is defined as an integral over the equilibrium distribution function
\begin{eqnarray}
\Sigma^\varepsilon(\tau,\tau^\prime) = \frac{8\pi^3}{\tau_{\rm eq}(\tau^\prime)} \int dw\, 
\int d^2p_\perp \,v \, G^{\rm eq}(\tau^\prime,\Delta h(\tau,\tau^\prime)+w, p_\perp).
\label{SigmaE}
\end{eqnarray}

%%%%%%%%%%%%%%%%%%%%%%%%%%%%%%%%%%%%%%%%%%%%%%%%%%%%%
\subsection{Longitudinal pressure}
%%%%%%%%%%%%%%%%%%%%%%%%%%%%%%%%%%%%%%%%%%%%%%%%%%%%%

In order to calculate the longitudinal pressure we use Eq.~(\ref{PL})
\begin{eqnarray}
P_\parallel(\tau) &=& \int dP \, \frac{w^2}{\tau^2} \left[ N_f \sum_{i} \left( G_{i}(x,p) + \bar{G}_{i}(x,p)\right) +\sum_{i>j} \left(
 \tilde{G}_{ij}(x,p) +  \tilde{G}_{ji}(x,p) \right)
 \right] \nonumber \\
 &=& \frac{2}{\tau^2} \int \, dw\,  \int d^2p_\perp \frac{w^2}{v} \left[ N_f \sum_{i}  G_{i}(x,p)  + \sum_{i>j}  \tilde{G}_{ij}(x,p)  \right]  .
  \label{PL1}
\end{eqnarray}
Using Eqs.~(\ref{formsol}) one finds
\begin{eqnarray}
P_\parallel(\tau) &=& \frac{N_f}{4\pi^3 \tau^2} \, 
\int_0^\tau d\tau^\prime D(\tau,\tau^\prime)
\sum_{i}
\left[2 \tau^\prime \Lambda_i^2(\tau^\prime)
C^-(\beta_i(\tau,\tau^\prime)) 
\frac{\Delta h_i^2(\tau,\tau^\prime)}{\sqrt{\Delta h_i^2(\tau,\tau^\prime)+\frac{\Lambda_i(\tau^\prime) \tau^2}{\pi}}}
+ \Sigma^{\parallel}_i(\tau,\tau^\prime) \right]
\nonumber \\
&& +\frac{1}{4\pi^3 \tau^2} \, 
\int_0^\tau d\tau^\prime D(\tau,\tau^\prime)
\sum_{i>j}
\left[2 \tau^\prime \Lambda_{ij}^2(\tau^\prime)
C^+(\beta_{ij}(\tau,\tau^\prime)) 
\frac{\Delta h_{ij}^2(\tau,\tau^\prime)}{\sqrt{\Delta h_{ij}^2(\tau,\tau^\prime)+\frac{\Lambda_{ij}(\tau^\prime) \tau^2}{\pi}}}
+ \Sigma^\parallel_{ij}(\tau,\tau^\prime) \right], \label{PL2}
\end{eqnarray}
where the functions $C^{\pm}(\beta)$ have been defined by Eq.~(\ref{Cpm}) and
\begin{eqnarray}
\Sigma^\parallel(\tau,\tau^\prime) = \frac{8\pi^3}{\tau_{\rm eq}(\tau^\prime)} \int dw\,
\int d^2p_\perp  \frac{w^2}{v}\, G^{\rm eq}(\tau^\prime,\Delta h(\tau,\tau^\prime)+w, p_\perp).
\label{SigmaL}
\end{eqnarray}

%%%%%%%%%%%%%%%%%%%%%%%%%%%%%%%%%%%%%%%%%%%%%%%%%%%%%
\subsection{Transverse pressure}
%%%%%%%%%%%%%%%%%%%%%%%%%%%%%%%%%%%%%%%%%%%%%%%%%%%%%

The transverse pressure is obtained from Eq.~(\ref{PT}),
\begin{eqnarray}
P_\perp(\tau) &=& \frac{1}{2} \int dP \, p_\perp^2 \left[ N_f \sum_{i} \left( G_{i}(x,p) + \bar{G}_{i}(x,p)\right) +\sum_{i>j} \left(
 \tilde{G}_{ij}(x,p) +  \tilde{G}_{ji}(x,p) \right)
 \right] \nonumber \\
 &=&  \int \, dw\,  \int d^2p_\perp \frac{p_\perp^2}{v}\, \left[ N_f \sum_{i}  G_{i}(x,p)  + \sum_{i>j}  \tilde{G}_{ij}(x,p)  \right]  .
  \label{PT1}
\end{eqnarray}
Using again the general forms of the solutions of the kinetic equations, see Eqs.~(\ref{formsol}), one finds
\begin{eqnarray}
P_\perp(\tau) &=& \frac{N_f}{4 \pi^3 \tau^2} \, 
\int_0^\tau d\tau^\prime D(\tau,\tau^\prime)
\sum_{i}
\left[
\frac{ \tau^\prime \Lambda_i^3(\tau^\prime)
E^-(\beta_i(\tau,\tau^\prime)) }{ \sqrt{\Delta h_i^2(\tau,\tau^\prime)+\frac{\Lambda_i(\tau^\prime) \tau^2}{\pi}}} \frac{\tau^2}{\pi}
+ \Sigma^{\perp}_i(\tau,\tau^\prime) \right]
\nonumber \\
&& +\frac{1}{4\pi^3 \tau^2} \, 
\int_0^\tau d\tau^\prime D(\tau,\tau^\prime)
\sum_{i>j}
\left[
\frac{ \tau^\prime \Lambda_{ij}^3(\tau^\prime)
E^+(\beta_{ij}(\tau,\tau^\prime)) }{\sqrt{\Delta h_{ij}^2(\tau,\tau^\prime)+\frac{\Lambda_{ij}(\tau^\prime) \tau^2}{\pi}}}
\frac{\tau^2}{\pi}
+ \Sigma^\perp_{ij}(\tau,\tau^\prime) \right], \label{PT2}
\end{eqnarray}
where
\begin{eqnarray}
\Sigma^\perp(\tau,\tau^\prime) = \frac{8\pi^3}{\tau_{\rm eq}(\tau^\prime)} \int dw\, 
\int d^2p_\perp \, \frac{p_\perp^2 \tau^2}{2 v} \,
G^{\rm eq}(\tau^\prime,\Delta h(\tau,\tau^\prime)+w, p_\perp)
\label{SigmaT}
\end{eqnarray}
and
\begin{eqnarray}
E^{\pm}(\beta) = \int_0^\infty
\,
\frac{d\xi \,\xi \, |\ln\left( 1\pm e^{-\xi}\right)|}{\sqrt{\beta^2+(1-\beta^2)\xi}}.
\label{Epm}
\end{eqnarray}
The functions (\ref{Cpm}), (\ref{Dpm}), and (\ref{Epm}) satisfy the constraint $D^\pm(\beta)= \beta^2 C^\pm(\beta) + (1-\beta^2) E^\pm(\beta)$.

%%%%%%%%%%%%%%%%%%%%%%%%%%%%%%%%%%%%%%%%%%%%%%%%%%%%%
\section{$\Sigma$ functions}
\label{app:sigmas}
%%%%%%%%%%%%%%%%%%%%%%%%%%%%%%%%%%%%%%%%%%%%%%%%%%%%%

The functions $\Sigma(\tau,\tau^\prime)$ defined by Eqs.~(\ref{SigmaJ}), (\ref{SigmaE}), (\ref{SigmaL}), and (\ref{SigmaT}) have a form of the integral
\begin{eqnarray}
\Sigma(\tau,\tau^\prime) &=& \frac{8\pi^3}{\tau_{\rm eq}(\tau^\prime)} \int dw\, 
\int d^2p_\perp \, X(\tau,w,p_\perp) \,
G^{\rm eq}(\tau^\prime,\Delta h(\tau,\tau^\prime)+w, p_\perp) \nonumber \\
&=& 4\pi \int\limits_{-\infty}^{+\infty} dw 
\int\limits_{0}^{\infty} p_\perp dp_\perp 
X(\tau,w,p_\perp) \,
\exp\left[-\frac{\sqrt{(\Delta h(\tau,\tau^\prime)+w)^2+p_\perp^2 \tau^{\prime \,2}}}{T(\tau^\prime) \tau^\prime} \right],
\label{Sigma1}
\end{eqnarray}
where the function $X(\tau,w,p_\perp)$ equals $w/v$, $v$, $w^2/v$, and $p_\perp^2 \tau^2/(2v)$, respectively (we recall that $v=\sqrt{w^2+p_\perp^2 \tau^2}$). Introducing the new integration variables $y=w+\Delta h$ and \mbox{$b=p_\perp/T(\tau^\prime)$}, and using the notation $T(\tau^\prime)=T^\prime$ we rewrite (\ref{Sigma1}) as 
\begin{eqnarray}
\Sigma
&=& 4\pi T^{\prime 2} \int\limits_{-\infty}^{+\infty} dy 
\int\limits_0^{\infty} b \,db \,
X(\tau,y-\Delta h,T^\prime b) \,
\exp\left[-\sqrt{\frac{y^2}{T^{\prime 2} \tau^{\prime 2}}+
b^2} \right].
\label{Sigma2}
\end{eqnarray}
In the next step we define $a=y/(T^\prime \tau^\prime)$ and obtain
\begin{eqnarray}
\Sigma
&=& 4\pi T^{\prime 3} \tau^\prime \int\limits_{-\infty}^{+\infty} da \int\limits_0^{\infty} b \,db \,
X\left(\tau,T^\prime \tau^\prime a -\Delta h,
T^\prime b\right) \,
\exp\left[-\sqrt{a^2+
b^2} \right].
\label{Sigma3}
\end{eqnarray}
Changing to the polar coordinates $a= r \cos\phi$ and $b=r\sin\phi$ we find
\begin{eqnarray}
\Sigma
&=& 4\pi T^{\prime 3} \tau^\prime  \int\limits_0^{\infty} e^{-r}\, r^2 \,dr \,
\int\limits_{0}^{\pi} d\phi \, \sin\phi \,
X\left(\tau,T^\prime \tau^\prime r \cos\phi -\Delta h,
T^\prime r\sin\phi\right) \,
.
\label{Sigma4}
\end{eqnarray}
In the case $\Sigma=\Sigma^J$ ($X=w/v$) 
\begin{eqnarray}
\Sigma^J(\tau,\tau^\prime) = 4 \pi T^3(\tau^\prime) \tau^\prime \int_0^\infty dr \, r^2 e^{-r} F^J \left(\frac{\Delta h(\tau,\tau^\prime)}{T(\tau^\prime)\tau^\prime r },\frac{\tau^\prime}{\tau} \right),
\label{SigmaJFJ}
\end{eqnarray}
where
\begin{eqnarray}
F^J(x,y) = y \int_0^\pi \frac{\sin\phi (\cos\phi-x)d\phi}{\sqrt{y^2 (\cos\phi-x)^2 + \sin^2\phi}}.
\label{FJ}
\end{eqnarray}
In the case $\Sigma=\Sigma^\varepsilon$ ($X=v$) 
\begin{eqnarray}
\Sigma^\varepsilon(\tau,\tau^\prime) = 4 \pi T^4(\tau^\prime) \tau^2 \int_0^\infty dr \, r^3 e^{-r} F^\varepsilon \left(\frac{\Delta h(\tau,\tau^\prime)}{T(\tau^\prime)\tau^\prime r },\frac{\tau^\prime}{\tau} \right),
\label{SigmaEFE}
\end{eqnarray}
where
\begin{eqnarray}
F^\varepsilon(x,y) = y \int_0^\pi \sin\phi \,
\sqrt{y^2 (\cos\phi-x)^2 + \sin^2\phi} \,d\phi.
\label{FE}
\end{eqnarray}
In the case $\Sigma=\Sigma^\parallel$ ($X=w^2/v$) 
\begin{eqnarray}
\Sigma^\parallel(\tau,\tau^\prime) = 4 \pi T^4(\tau^\prime) \tau^2 \int_0^\infty dr \, r^3 e^{-r} F^\parallel \left(\frac{\Delta h(\tau,\tau^\prime)}{T(\tau^\prime)\tau^\prime r },\frac{\tau^\prime}{\tau} \right),
\label{SigmaLFL}
\end{eqnarray}
where
\begin{eqnarray}
F^\parallel(x,y) = y^3 \int_0^\pi \frac{\sin\phi (\cos\phi-x)^2  d\phi}{\sqrt{y^2 (\cos\phi-x)^2 + \sin^2\phi}}.
\label{FL}
\end{eqnarray}
We note that the definition of the variable $v$, see Eq.~(\ref{binvv2}), implies that $\Sigma^\varepsilon = \Sigma^\parallel + 2 \Sigma^\perp$. Hence, the function $\Sigma^\perp$ may be expressed in terms of $\Sigma^\varepsilon$ and $\Sigma^\parallel$. We also note that the integrals (\ref{FJ}), (\ref{FE}), and (\ref{FL}) are analytic.

\end{document}